\documentclass[graybox, nosecnum]{svmult}

\usepackage{textgreek}
\usepackage{mathptmx}       % selects Times Roman as basic font
\usepackage{helvet}         % selects Helvetica as sans-serif font
\usepackage{courier}        % selects Courier as typewriter font
\usepackage{type1cm}        % activate if the above 3 fonts are
\usepackage{makeidx}         % allows index generation
\usepackage{graphicx}        % standard LaTeX graphics tool
\usepackage{multicol}        % used for the two-column index
\usepackage[bottom]{footmisc}% places footnotes at page bottom
\usepackage{hyperref}        %for hyperlinks
\usepackage{soul}            % for high-lighting of text https://www.overleaf.com/project/6152f7f1fe2c89b11547be41
\usepackage{float} % for forcing figures in certain places
\usepackage{ulem} % for strike out text
\hypersetup{colorlinks=true,urlcolor=blue}
\usepackage[square,numbers]{natbib}
\usepackage{caption}
\usepackage{subcaption}
\makeindex             % used for the subject index
\newcommand{\lsi}{LS I$+61^{\circ}303$}

\begin{document}
%\tableofcontents{}
\title*{The Very Energetic Radiation Imaging Telescope Array System (VERITAS)}
%\titlerunning{VERITAS} 
\titlerunning{The Very Energetic Radiation Imaging Telescope Array System (VERITAS)}
\author{David Hanna and Reshmi Mukherjee %\thanks{corresponding author}
}
\authorrunning{D. Hanna and R. Mukherjee}
%\authorrunning{Mukherjee \& Hanna}
%Use \authorrunning{Short Title} for an abbreviated version of
\institute{David Hanna \at Physics Department, McGill University, Montreal, QC H3A 2T8, Canada, \email{hanna@physics.mcgill.ca}
\and Reshmi Mukherjee \at Department of Physics \& Astronomy, Barnard College, Columbia University, New York, NY 10027, USA, \email{rm34@columbia.edu}}
\maketitle
\section{Abstract}
The VERITAS observatory, located in southern Arizona, is engaged in an exploration of the gamma-ray sky at energies above 85 GeV. Observations of Galactic and extragalactic sources in the TeV band provide clues to the highly energetic processes occurring
in these objects, and could provide indirect evidence for the origin of cosmic rays and the sites of particle acceleration in the Universe. In this chapter, we describe the VERITAS telescopes and their operation, as well as analysis procedures, and present results from scientific observations, which include extragalactic science, Galactic physics, and studies of fundamental physics and cosmology.

\section{Keywords} 
Astrophysics - Instrumentation and Methods for Astrophysics; Astrophysics - High Energy Astrophysical Phenomena
Astroparticle physics; Cherenkov
telescopes

\section{Introduction}

The VERITAS instrument is a four-telescope array located on the lower slopes of Mount Hopkins, approximately 60 km south of Tuscon, Arizona (31°40'N, 110°57'W, 1270 m asl). It is one of three \lq third-generation\rq $~$  gamma-ray observatories built during the early 2000s. The other two are H.E.S.S. and MAGIC, both described elsewhere in this volume. 

\begin{figure}[b]
\centering
\includegraphics[width=0.99\textwidth]{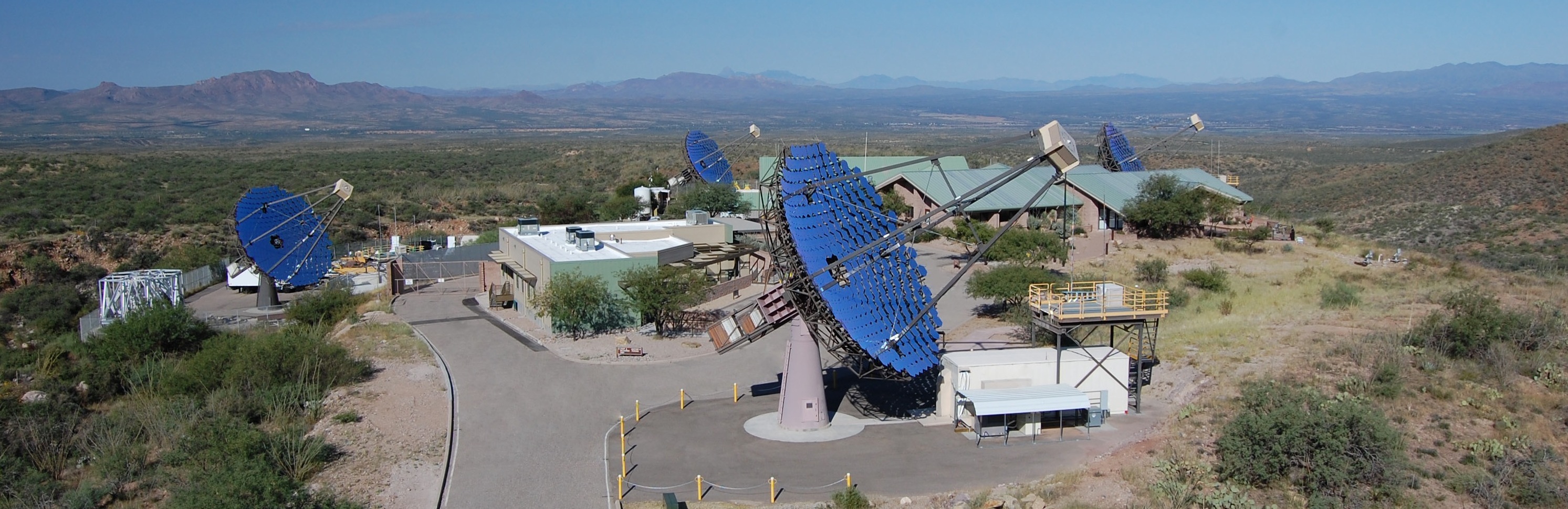}
\caption{{\small  A photograph of the VERITAS telescopes located at the Fred Lawrence Whipple Observatory. (Source: Steve Criswell and the VERITAS Collaboration.)}}
\label{fig:veritas_and_building}
\end{figure}

The idea of an array of imaging atmospheric-Cherenkov telescopes (IACTs) had been around for many years, even before the first successful detection of an astrophysical source using an IACT. The detection of the Crab Nebula by the Whipple group in 1999~\citep{crab-10m} was a major step forward, giving rise to a series of workshops starting in Palaiseau in 1992~\cite{fleury}. During this decade an enlarged group based on the original Whipple collaboration proposed an array of seven 12-m telescopes to be built on Mount Hopkins. Following a lengthy process of descoping to address fiscal realities,  and a change of location motivated by concerns of some of the local public, a four-telescope array was eventually built at the Fred Lawrence Whipple Observatory base camp. A partially instrumented telescope, serving as a prototype~\cite{holder}, was installed in 2004 and the complete array saw first light in 2007 (Fig.~\ref{fig:veritas_and_building})~\cite{hanna-1}.

\section{Telescopes}

\begin{figure}[t]
\centering
\includegraphics[width=0.99\textwidth]{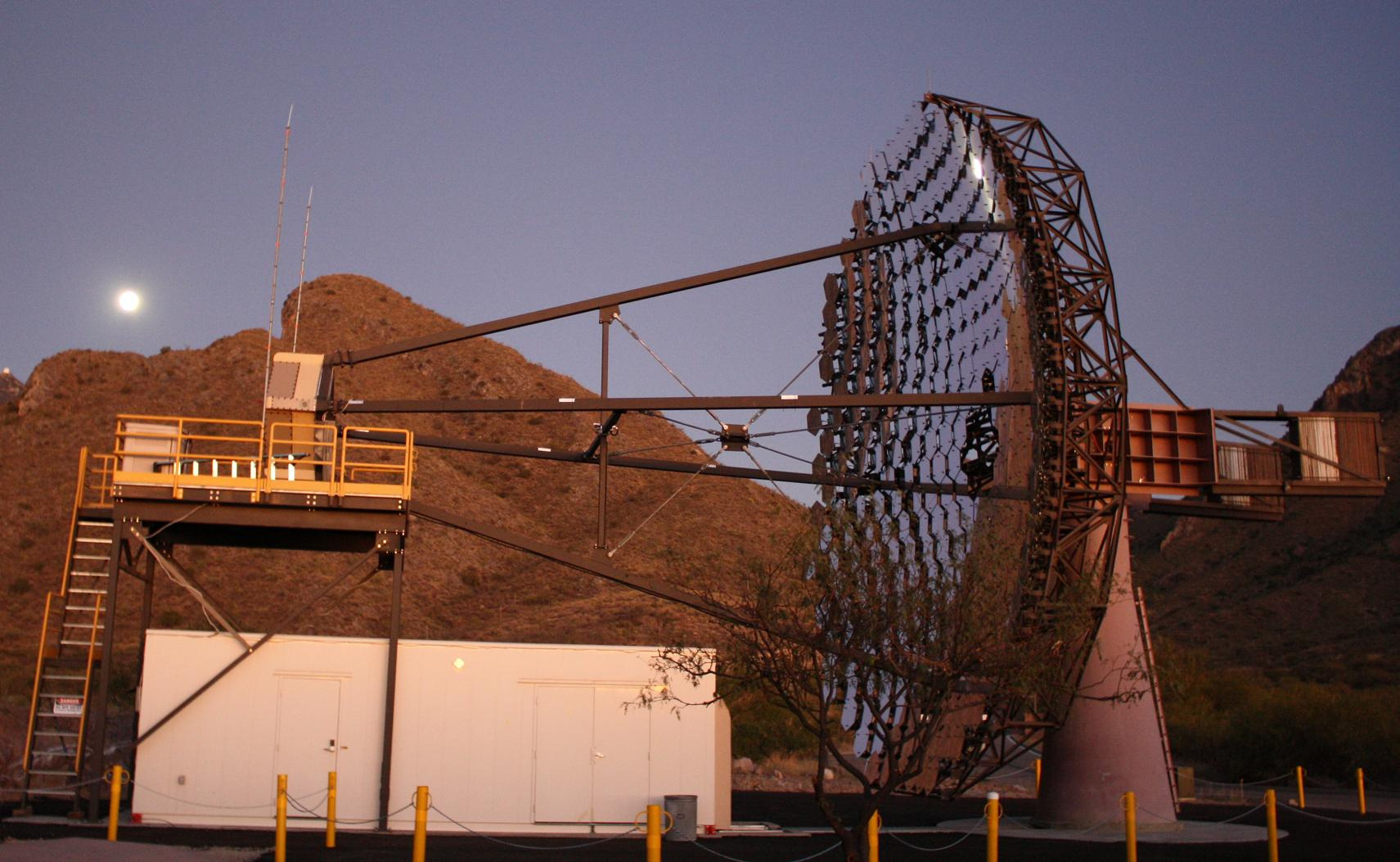}
\caption{{\small  A photograph of one of the VERITAS telescopes in stow position. The focal-plane box containing the PMT-based camera can be seen at the left
side, along with a service platform. The building directly below houses the digitizing electronics and high-voltage supplies. 
Cables for these exit the right end of the building and are routed up through the conical alt-az positioner and out to the camera through the four hollow beams that make up the quadrupod. The focal-plane box is connected to the reflector structure by the quadrupod. The reflector comprises 345 hexagonal mirror facets installed on a tubular steel space frame. Counterweights made from steel plates, for fine-tuning, can be seen at the right. (Photo: S. Griffin.)}}
\label{fig:t1-photo}
\end{figure}

All four VERITAS telescopes are nominally identical and follow closely the design of the original Whipple 10 m telescope~\cite{kildea}. 
Each telescope has a 12-m-diameter reflector that collects and directs Cherenkov light onto a camera comprising 499 pixels, located at the focal point 12 m from the mirror. The reflector/camera combination is mounted on an alt-az drive capable of slew speeds of the order of a degree per second (Fig.~\ref{fig:t1-photo}).

\subsection{Reflectors}

The reflectors follow the classic Davies-Cotton design~\cite{davies} and consist of 345 identical mirror facets mounted on a tubular-steel optical support system (OSS). Each facet~\cite{roache} has a hexagonal shape to minimize dead space, with a distance of 610 mm between opposite vertices. They are made from soda lime float glass, 11.5 mm thick, slumped and polished to give a 24-m focal length, and are coated with a 180 nm layer of aluminum at an on-site facility maintained by the VERITAS collaboration. The facets are front-surfaced to avoid absorption of the large ultraviolet fraction of Cherenkov light. This means that they are vulnerable to weathering effects; the aluminum is anodized to a depth of 80 nm to combat this.

100 spare mirror facets are used in a maintenance program whereby a group of 100 weathered facets is replaced by freshly coated facets. These swaps occur about three times per year, limited by the throughput of the coating facility. Bench-top monitoring of the reflectivity of a sample set of facets from all telescopes has been in place since the start-up of VERITAS observations in 2007 and a system to monitor the effective reflectivity of the entire reflector was installed on each telescope in 2014. This device~\cite{archambault} uses a CCD camera mounted near the center of the reflector in such a way as to simultaneously view a bright star and its reflection, as seen on a Spectralon screen temporarily mounted in the focal plane of the telescope. The ratio of the intensities of the star and the reflection are used in calculating the effective reflectivity.

Each facet is secured to the OSS with three gimbal-based devices that allow the precise direction of the facet to be set in order to achieve the best overall optical point-spread function for the full reflector. Originally, alignment was carried out using laser measurements made with the telescope at stow position, with corrections for flexing of the OSS when deployed for observations~\cite{toner}. This was replaced in 2010 with a new system~\cite{mccann} that employs a CCD camera temporarily mounted at the telescope focal point and looking back towards the mirror facets. Images are recorded during a raster scan of a bright star and the brightness of each facet as a function of scan position is used to compute mechanical adjustments that are made the following day. The mounting system is sufficiently stable that use of the device is only necessary after facet-swap events. The resulting optical PSF (80\% containment radius) is less than 0.075$^\circ$ at all zenith angles. 

\subsection{Cameras}

At the focal point of each reflector dish is a camera made of 499 photomultiplier tubes (PMTs)~\cite{nagai}. The PMTs have a nominal diameter of 26 mm and are mounted in an approximately circular close-packed hexagonal arrangement. Each is equipped with a truncated Winston cone~\cite{winston} to increase its effective area, reduce the dead space between PMTs, and restrict the angular acceptance of the PMT to light coming from the reflector. This, along with the quantum efficiency, which peaks near 300 nm, optimizes performance for Cherenkov light from air showers while reducing the sensitivity to night-sky background (NSB). The optical geometry is such that each PMT has a field-of-view (FOV) of 0.15$^\circ$ and the entire camera has an FOV of 3.5$^\circ.$

Initially, the PMTs were 29 mm Photonis XP2970 PMTs, with a peak quantum efficiency (QE) of approximately 25\% at 320 nm.  These were replaced, during the summer of 2012, with Hamamatsu R10560-100-20 PMTs that have peak QE of 35\% at 300 nm~\cite{kieda}.They are run with a gain of approximately $2 \times 10^5$ with high-voltage (HV) values near 1 kV. 

\begin{figure}[t]
     \centering
     \begin{subfigure}[b]{0.45\textwidth}
         \centering
         \includegraphics[width=\textwidth]{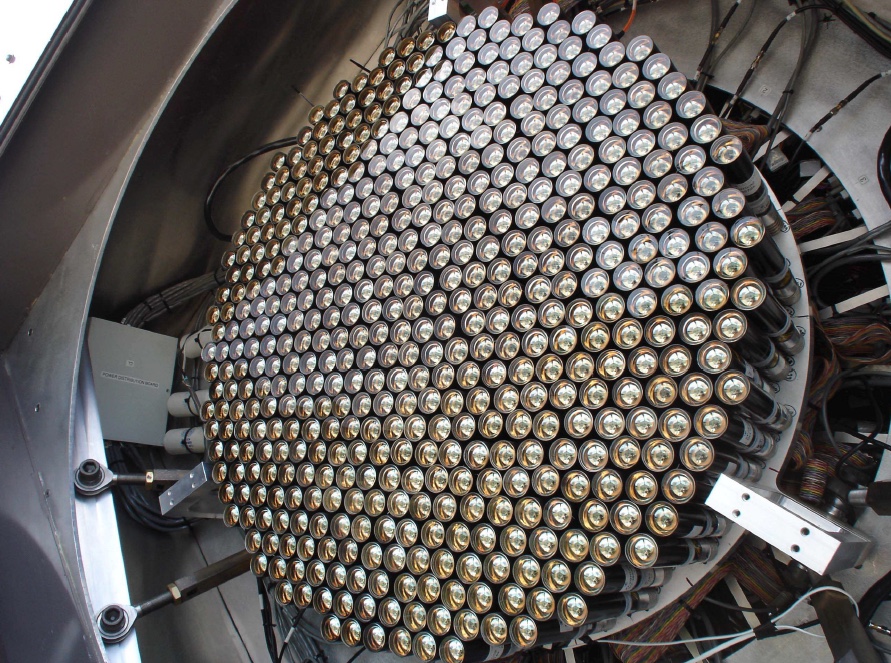}
     \end{subfigure}
     \begin{subfigure}[b]{0.45\textwidth}
         \centering
         \includegraphics[width=\textwidth]{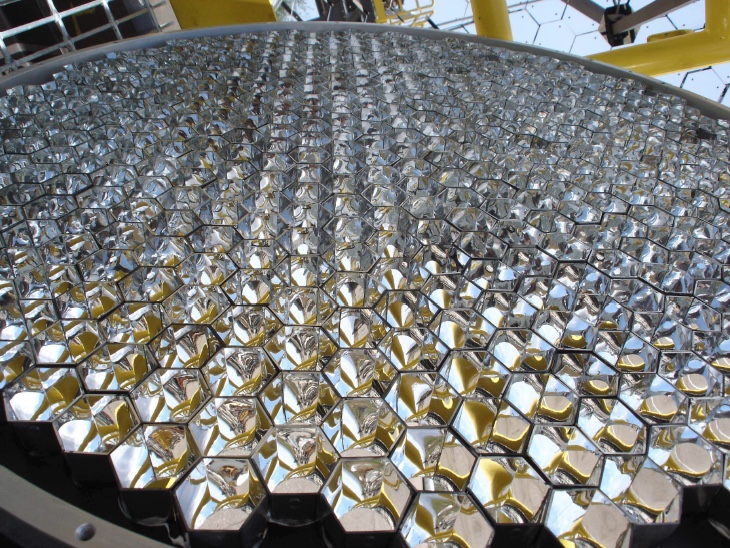}
     \end{subfigure}
        \caption{\small Left: Camera, before installation of the Winston cones, showing the close-packed-hexagonal layout of the PMTs. Right: The truncated Winston cones used to reduce the effect of inter-PMT deadspace and define the angular acceptance of the camera. (Source: VERITAS Collaboration.)}
    \label{fig:camera-cones}
\end{figure}

Each PMT is integrated into a pixel. A pixel comprises the PMT and a resistive voltage divider to bias the dynodes. Also included are a mu-metal shield and a preamp which provides a nominal gain of 6.6. An on-board monitor produces a DC signal proportional to the average photocurrent. RG59 75-\textOmega $~$cables are used to transport the PMT pulses to the digitizing electronics located in an electronics building adjacent to each telescope. HV supplies (CAEN) to power the PMTs are also located there. The camera is housed in a metal box stood off from the reflector by a steel quadrupod, the arms of which are hollow and serve as conduits for the cables (Fig.~\ref{fig:camera-cones}).

\subsection{Electronics and Data Acquisition}

Signals from the PMTs are digitized using flash analog-to-digital converters (fADCs) purpose-built by members of the VERITAS collaboration. Each fADC produces an eight-bit measurement of the signal every two nanoseconds. The sampling rate is well-matched to the pulse length; a 16-sample window captures the entire pulse and with enough granularity to achieve ns-level timing precision. Additional amplitude dynamic range is provided by a high/low gain capability. Two copies of the pulse are produced, one of which is delayed and attenuated by a factor of approximately 6. If the nominal (high-gain) pulse exceeds a saturation threshold, a switch is closed and the attenuated (low-gain) pulse is digitized. The fADC modules also host a constant-fraction discriminator (CFD) for each signal. The fADCs are installed on circuit boards, 10 channels per board, located in custom VME crates, four per telescope, in water-cooled racks. The crates are controlled by local single-board computers. 

The digitizers run continuously, each writing to a circular ``ring'' buffer that can accommodate 32 $\mu$s of data and serves as a delay pipeline while the trigger is formed. VERITAS employs a three-level trigger, starting (L1) with the CFDs to define which PMTs have pulses exceeding a user-defined threshold. The level is determined using rate-vs-threshold runs which show where array-level triggers from air showers exceed those arising from NSB.  The second level (L2) is a pattern trigger which requires clusters of \lq struck\rq $~$PMTs. It is implemented locally on each telescope and a typical requirement is to have three adjacent struck PMTs with firing times within 6 ns of each other. The L2 trigger was upgraded in 2012~\cite{zitzer} using electronics based on field-programmable gate arrays (FPGAs) to enable higher rates and more flexibility. The final (L3) trigger is an array trigger~\cite{weinstein}. It requires a given number of telescopes (typically two) to satisfy the L2 requirements within a given time window (typically 50 ns). The window is shorter than that required to accommodate the inter-telescope delays that arise from purely geometrical effects due to the angle of the incoming burst of Cherenkov light. These delays are calculated in real time and compensated for by applying dynamic delays to the L2 signals before they are used in the L3 trigger. 

When an L3 trigger is generated, the digitizers from all four telescopes are read out. Using a dynamic look-back time, a 32 ns (16-sample) segment of each ring buffer is extracted by the local computer in each crate and written to an 8 MB buffer. Full buffers are sent to a fifth computer which acts as an event builder, writing this information, along with a timestamp, to a local disk. A central computer, the harvester, combines event files from the four telescopes into a single array file and compresses it. Readout of a complete event takes about 400 $\mu$s, so deadtime is about 12\% for the 300 Hz trigger rates that VERITAS typically experiences.  Data runs are usually 30 minutes in length and the resulting files are approximately 10 GB in size. They are archived off-site.

\subsection{Diagnostic and Monitoring Systems}

To maintain the high quality of data recorded by the VERITAS telescopes, various systems have been developed. They will be described in the following paragraphs.

Since IACTs rely on Cherenkov light coming from the entire air shower, clarity of the atmosphere up to altitudes higher than 10 km is an important consideration. An obvious problem is the existence of local clouds which can attenuate light from the showers and diffusively reflect nearby sources of light pollution. Observers can assign a quality factor to the data based on their subjective impressions but two, more quantitative, instruments are also used. 

The first is a pair of far-infrared radiation (FIR) pyrometers. Each is mounted on the edge of a telescope reflector such that it views the same part of the sky as that being tracked during observations. Clouds, whether visible to the naked eye or not, show up as an increase in the temperature measured by the pyrometers, which is recorded as part of the data stream. The FIR instruments have been used since the beginning of VERITAS operations and have been supplemented since 2012 by a commercially available LIDAR system (Vaisala CL51) that is aimed at the zenith and uses backscattered 910 nm light emitted from an InGaAS diode laser to detect and measure clouds up to an altitude of 13 km. Data from this device are also recorded so as to enable off-line weather-quality cuts.

The alt-az positioners used for the telescopes have pointing accuracy of order 0.01$^\circ$. To improve on this, two CCD cameras are mounted on each telescope. One, the sky camera, is aligned with the telescope's optical axis and is pointed at the sky. The other, the focal-plane camera, views the plane directly in front of the PMTs. A quartet of red LEDs attached at the edges of the focal plane serve as calibration markers. During monthly calibration runs, a thin white screen is attached in front of the focal plane and images from the two CCD cameras are acquired while the telescope is successively pointed to a number of bright stars covering a large set of pointing directions. Using the image of the star in the sky camera and the image of its reflection in the focal-plane camera (along with the LED images) allows one to construct a set of correction factors that can be applied to the pointing direction. During normal data runs, images from the two cameras are analyzed in real time and the coordinates of the stars in the sky-camera FOV and the LEDs in the focal-plane camera are recorded. The associated pointing corrections are calculated off-line. These corrections reduce absolute pointing errors to less than 10 arcseconds.

To measure and monitor the gains of the PMTs, each telescope is equipped with an LED-based flasher system~\cite{hanna-2}. Each flasher contains a number, {\it N}, of UV LEDs (peak wavelength at approximately 370 nm) that are pulsed in such a way that the flasher's light intensity rises in steps from 0 to {\it N} in a repeating cycle. {\it N} was  initially 7 but was increased to 15 after the upgrade to new PMTs in 2012. The flashers are used every night when observations are being carried out; a single two-minute run with the flashers pulsed at 300 Hz provides enough data for a gain estimate for every PMT. During nights with partial moonlight, VERITAS is run with reduced gains~\cite{archambault-2}, so a second flasher run is sometimes necessary. Several times per year the flashers are used in determining the absolute gain scale by measuring the single-photoelectron (spe) peak seen when the PMTs are illuminated at low light levels. To attenuate the flasher light, as well as NSB, so that the spe signal is not overwhelmed, a thin aluminum plate with a hole of 3-mm diameter in front of every PMT is installed in front of the Winston cones~\cite{hanna-3} to act as a neutral-density filter. The flashers are also periodically used for measuring the high/low gain ratio for each channel in the fADC system. 

\subsection{Telescope Positions}

\begin{figure}[t]
\centering
\includegraphics[width=0.99\textwidth]{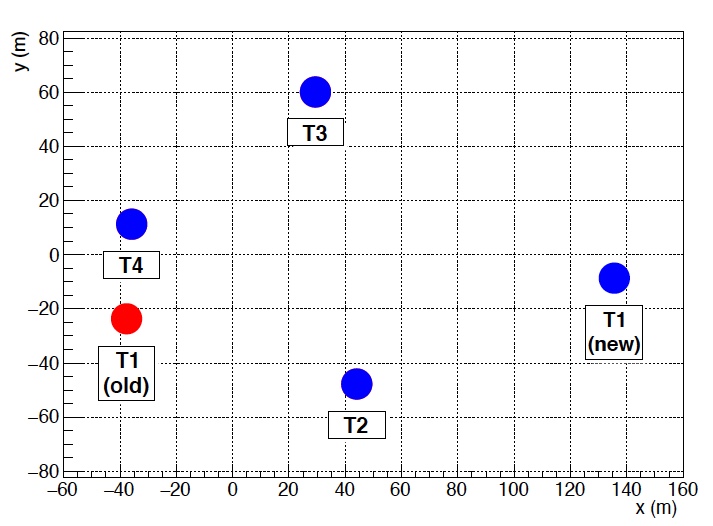}
\caption{{\small Layout of the VERITAS telescopes with the y axis pointing north. Telescope 1, originally installed in the \lq old\rq$~$position was relocated to the \lq new\rq$~$position during the summer shutdown of 2009 in order to rationalize the layout and improve the array's sensitivity. (Diagram: S. Griffin.)}}
\label{layout}
\end{figure}

The relative positions of the VERITAS telescopes are shown in Fig.~\ref{layout}. The original VERITAS four-telescope design called for a Y-shaped layout with a
telescope at each vertex of an equilateral triangle with 140 m sides and one telescope at the center. Difficulties in obtaining construction permission at the site originally planned forced a relocation to the Whipple Observatory base camp, where the first telescope, T1, had been constructed as a prototype~\cite{holder}. Existing base-camp infrastructure (see Fig.~\ref{fig:veritas_and_building}) constrained the location of the other telescopes. Shortly after commissioning the complete array, it was decided to improve the layout by relocating T1 to its present position. The move was made during the summer shutdown of 2009.

\section{Performance}

VERITAS operations are carried out each year from September through the following June. During the summer, unstable weather conditions associated with the annual monsoon are unsuitable for astronomy; to avoid damage to electronics from lightning strikes, all cables running from the PMTs are disconnected. At other times, observations are made during dark and partially moonlit nights. In a typical season 1000 hours of observations are made, with 75\% under fully dark skies. More than 95\% of the time, all four telescopes in the array have participated.

The nominal performance characteristics of the instrument are:
\begin{itemize}
    \item energy range: 85 GeV to $>$ 30 TeV (spectral reconstruction starts at 100 GeV)
    \item energy resolution: 17\% at 1 TeV
    \item peak effective area: 100 000 m$^2$
    \item angular resolution: 0.08$^\circ$ at 1 TeV, 0.13$^\circ$ at 200 GeV (68\% containment radius)
    \item source location accuracy: 50 arcseconds
    \item point source sensitivity: 1\% Crab Nebula flux in $<$ 25 hours
\end{itemize}

\section{Components for Ancillary Science}

Cherenkov telescopes have some of the largest light-collecting areas in astronomy. Their optics are not as refined as those found in more conventional telescopes but for certain applications this is not a handicap. Having two or more telescopes in the array and with the capabilities of high-speed data acquisition, they are well suited for intensity interferometry and asteroid-occultation measurements.

Stellar intensity interferometry was developed by Hanbury Brown and Twiss~\cite{sii-1} and used to measure the angular diameter of bright stars. The technique has been used at VERITAS since 2019 and observations of over 40 stars,  with $m_v < 3.8$ and angular diameters from 0.4 milliarcsecond (mas) to 1.2 mas, have been made during the bright-moon periods when gamma-ray observations are not possible~\cite{sii-2}. 

The apparatus constructed for each camera, for these measurements, consists of a PMT of the type used in the VERITAS cameras (Hamamatsu R10560 super-bialkali) and a Semrock FF01-420/5-2 narrow-band filter (5 nm width at 420 nm). These are mounted on an aluminum plate equipped with a 45-degree mirror such that the assembly can be attached in front of the Winston cones with the filter/PMT combination being virtually on-axis~\cite{sii-3}. This means that no changes to the main camera are required and the switch to optical measurements requires very little time. 

The signal from the PMT is sent through a local high-speed preamplifier and then through 45 m of coaxial cable to a 250 Msps 12-bit DC-coupled digitizer located in the adjacent electronics building. The data are truncated to 8 bits before being streamed onto disk.  To achieve high bandwidth and consequent high signal-to-noise ratio, the digitizers are synchronized using a White Rabbit~\cite{rabbit} timing system and single-mode optical fiber links.

Another application that makes use of the large areas of the VERITAS telescope mirrors is that of high-time-resolution precision photometry measurements. By making use of chance occultations of stars by asteroids the stellar diameters can be determined with precision better than 0.1 mas. This is done by fitting the diffraction fringes seen on the edges of the shadow cast by the obscuring object, recorded as the shadow passes through the telescope's field of view. Lunar occultation has been used in this way for many years~\cite{morbey}, but use of asteroids had not met with success until VERITAS measurements were made~\cite{asteroid}. 

To perform the measurements, 16 central pixels in each camera were equipped with electronics components to monitor their DC light levels. A 14-bit commercially available voltage datalogger with sampling rates of up to 4.8 kHz was used and implemented in parallel with the standard data-acquisition system so as not to disrupt gamma-ray observations. An upgrade to the VERITAS fADC system has been funded and this will make it possible to record DC light levels on all the PMTs in each camera as part of the standard data stream, opening up further possibilities in this area.

\section{The Scientific Program of VERITAS}

The scientific program of VERITAS has been guided by the goal of trying to understand the physics of cosmic accelerators and multimessenger sources. Very-high-energy (VHE, energy $E > 100$ GeV) gamma-ray observations offer indirect methods for studying the highest-energy cosmic rays in the Universe. 
Since cosmic rays are charged particles they are deflected by interstellar and intergalactic magnetic fields, and (except at the very highest energies) cannot be traced back to their sources. Neutral messengers such as gamma rays and neutrinos, that may be produced at the same sites, are perhaps the best probes for locating particle accelerators. In astrophysical sources, gamma rays can only be produced via nonthermal processes such as in the acceleration of relativistic charged particles, for example by a magnetic field or a plasma shock. In addition, gamma rays may be used to probe intergalactic space and and explore diffuse radiation fields. Gamma rays provide a crucial window on the cosmic electromagnetic spectrum, and gamma-ray data from astrophysical sources may be used to explore fundamental physics topics. 

\begin{figure}[t]
\centering
\includegraphics[width=0.99\textwidth]{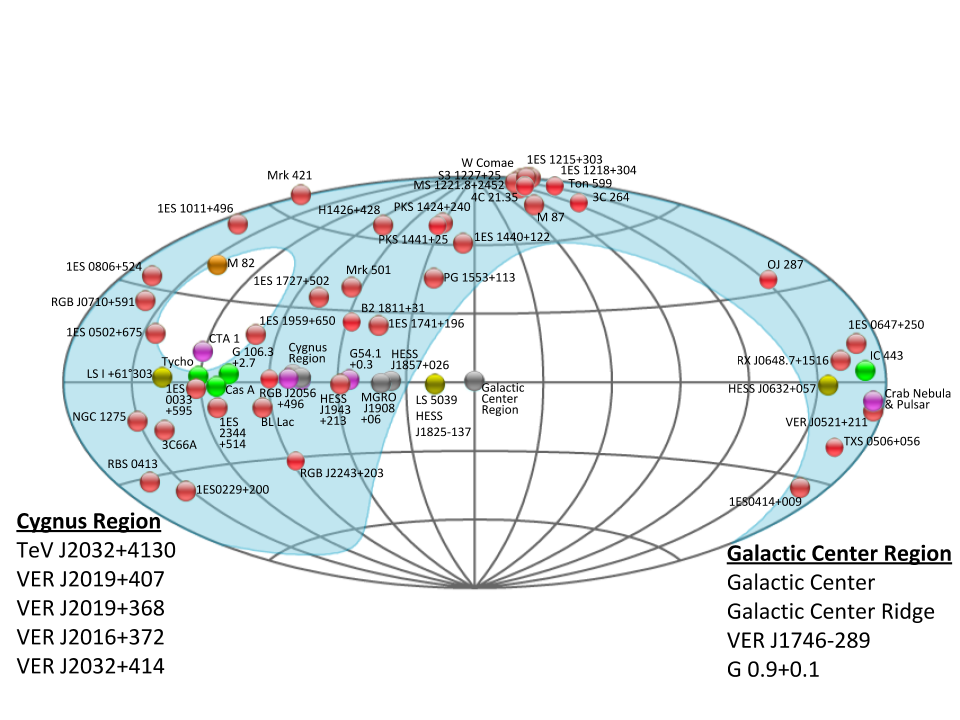}
\caption{{\small  A sky map showing the sources detected by VERITAS as of November 2021. Different source types are indicated as follows: blazars (red), pulsar wind nebulae (magenta), binary star systems (yellow), supernova shells (green), starburst galaxies (orange), and dark/unidentified (grey). VERITAS views the northern sky --  the portion of the sky visible from the observatory is demarcated by the shaded area. (Figure
generated using TeVCat~\cite{tevcat}.)}}
\label{fig:veritas_skymap}
\end{figure}

VERITAS observations have been ongoing since ``first light'' in 2007~\cite{hanna-1} and data have been collected for nearly sixteen years. Figure~\ref{fig:veritas_skymap} shows the sky map of sources detected by VERITAS, with the source count currently at 65. The TeV sky is diverse, with eight different source classes detected by VERITAS, including blazars (BL Lacertae objects and flat-spectrum radio quasars), radio galaxies, pulsar wind nebulae, pulsars, supernova shells, binary star systems and starburst galaxies. "Unidentified" sources are detected only at TeV energies and have no known counterparts at other energies. In the next sections, a few key findings from the VERITAS observing program are summarized. 

\subsection{Extragalactic Source Studies}

The majority of the extragalactic sources detected by VERITAS are blazars, objects powered by supermassive black holes (SMBHs), belonging to a class of active galactic nuclei (AGN) that have ultra-relativistic jets (outflows of particles) pointed close to our line of sight~\cite{urry}. Blazars constitute the largest fraction of extragalactic TeV sources and these VHE 
blazars are the most extreme form of AGN. 
Almost all gamma-ray-emitting AGN are ``radio loud'' or bright radio sources, with flat spectra at radio energies, where ``flat'' signifies the relationship between energy flux as a function of energy in a logarithmic scale, thus indicating a lack of spectral cutoff. These sources are characterized by emissions that include high radio and optical polarization, strong
variability at all wavelengths, extremely high luminosities, apparent superluminal velocities of compact radio cores, and nonthermal continuum spectra~\cite{urry}. 
Studies of gamma-ray emission from blazar jets are key to understanding the particle composition of the jets (whether leptonic or hadronic), the nature of particle-acceleration mechanisms, the energy content of the jet, and how VHE blazars may contribute to the highest-energy cosmic ray and neutrino fluxes. 
Gamma rays with energies greater than 10 TeV are routinely detected from AGNs. These are likely born in processes involving charged particles that have been accelerated to even higher energies. If those particles are protons, it is probable that neutrinos with similar energies will also be produced. Accelerated protons that escape from AGNs could constitute all, or a substantial part, of the cosmic ray flux at energies higher than 10 TeV.

In addition to blazars, the other main category of gamma-ray-emitting radio-loud AGN that have also been detected at TeV energies are radio galaxies, sources with powerful radio jets. 

\subsubsection{The VERITAS Blazar Sample} 

\begin{figure} [b!]
\centering
\vskip -0.1in
\includegraphics[height=0.31\textwidth]
 {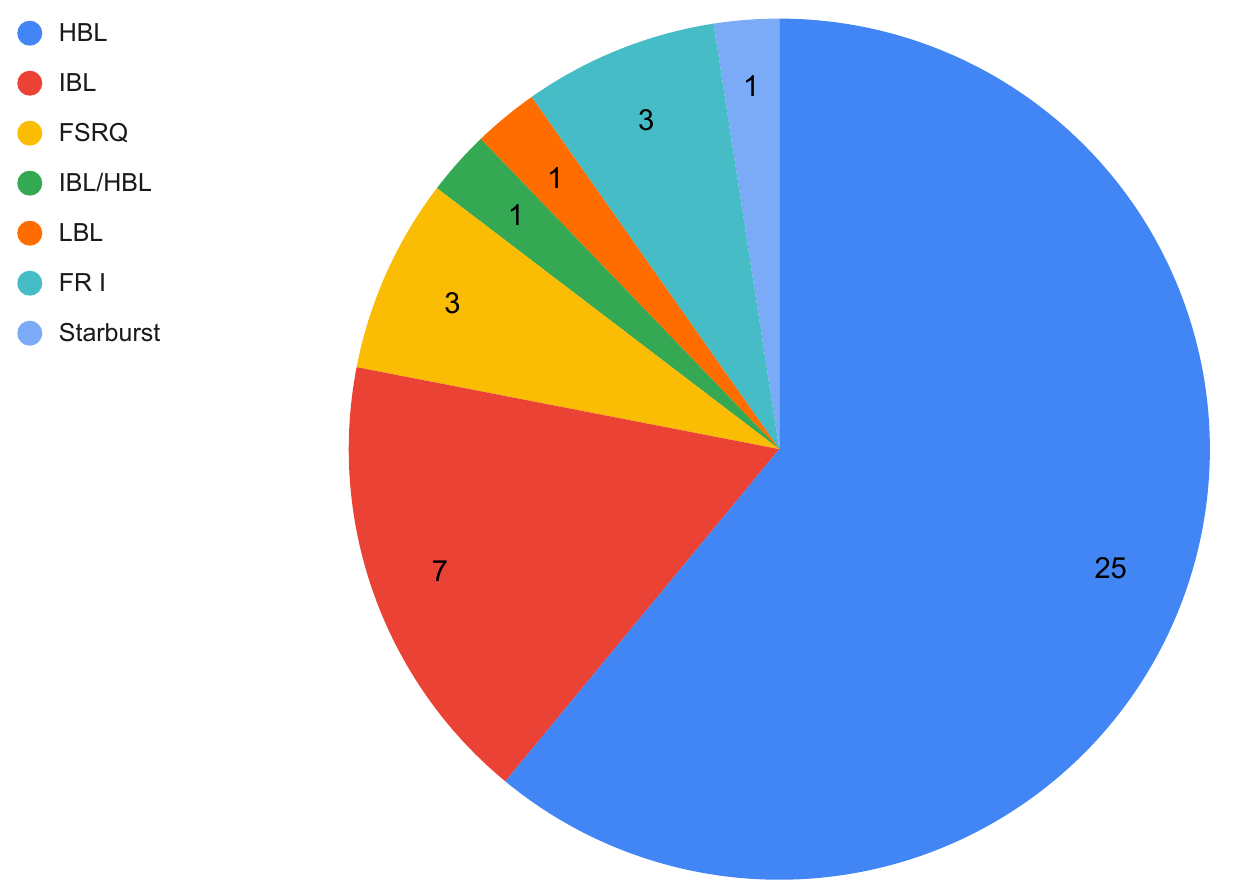}
\includegraphics[height=0.31\textwidth]{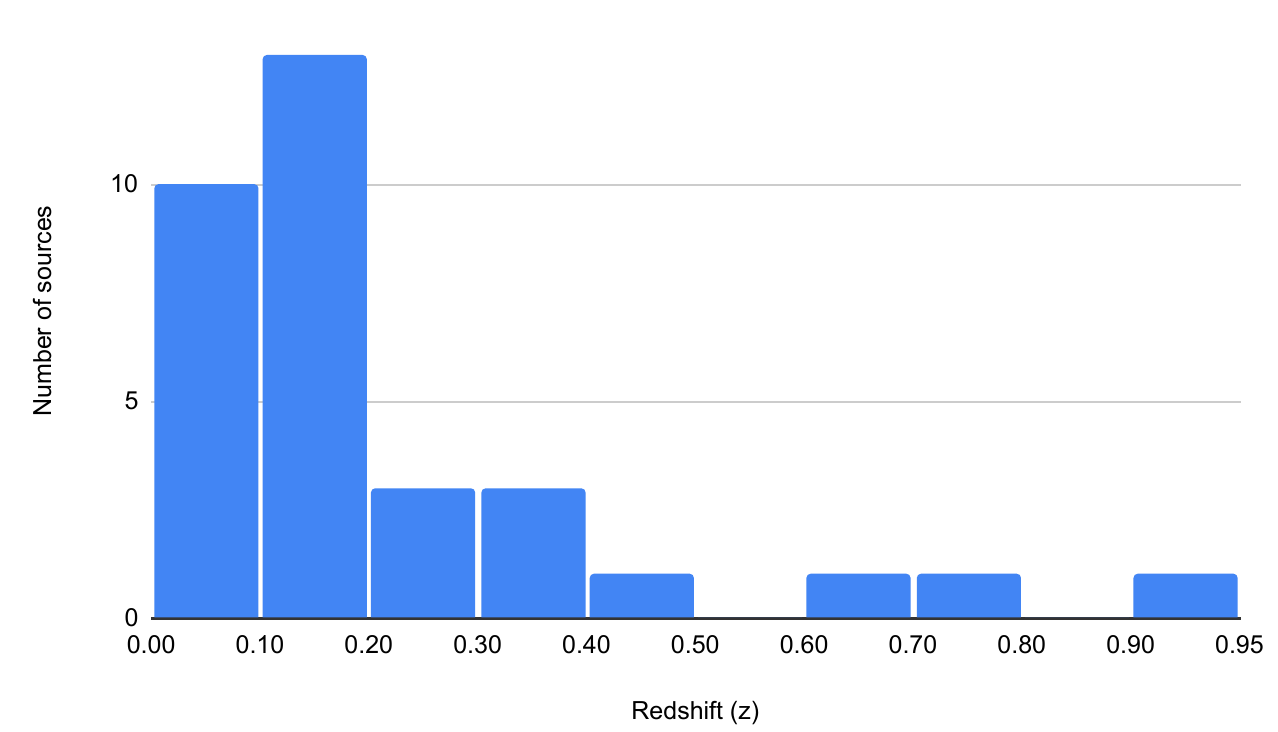}%\vskip -0.1in
  \caption{ \small (a) Left: Counts of AGN by type detected by VERITAS. 
(b) Right: Redshift distribution of the AGN detected by VERITAS. Data for both figures are as of October 2021. 
}
\label{fig:veritas_redshift_agn}
\end{figure}

Blazars may be divided into the subclasses of low-luminosity BL Lac objects (BL Lacs) and high-luminosity FSRQs (flat-spectrum radio quasars). In the current understanding,  blazars have an accretion disk that surrounds a SMBH at the center, ringed by a dusty molecular torus~\cite{urry}, with  relativistic jets along the rotation axis of the torus. BL Lac objects are characterized by low accretion rates whereas FSRQs exhibit a high accretion rate, and these characteristics influence the spectral energy distributions for these objects. 
Based on physical properties, BL Lacs are traditionally subclassified into HBLs, IBLs and LBLs (high-, intermediate- and low-frequency-peaked BL Lacs), according to the peak frequency of their synchrotron emission, following the classification in~\cite{padovani1995}.
%The VERITAS blazar sample include xx BL Lac objects, yy FSRQs, and zz radio galaxies, with detections up to a redshift of 1. 
Figure~\ref{fig:veritas_redshift_agn} shows the distribution of the different categories of extragalactic sources detected by VERITAS. Not surprisingly, HBLs dominate the blazar population. The most famous example of an HBL is Markarian 421 (Mrk 421), the first extragalactic object to be detected in the VHE band~\cite{punch1992}.
%  (also see~\cite{whipple_mrk421} for a 14-year study of the source by the Whipple Observatory). 

Only three FSRQs have been detected, so far, by VERITAS, despite the higher bolometric luminosities of these objects~\cite{ari_fsrq_2021}. In fact, FSRQs are typically detected by VERITAS only in high flaring states. In 2020, VERITAS began a survey of FSRQs in a systematic unbiased way~\cite{sonal_patel_icrc2021}, which when completed, could offer clues to the duty cycles of FSRQs.  

Beyond an energy- and flux-dependent redshift, gamma rays from blazars are not detected on Earth due to gamma-gamma absorption of VHE photons in pair-production interactions with optical-infrared photons of the extragalactic background light (EBL). Figure~\ref{fig:veritas_redshift_agn} shows the redshift distribution of the AGN detected by VERITAS. About 90\% are at a redshift less than $0.5$. The most distant blazar detected by VERITAS is the FSRQ PKS 1441+25, at a redshift of 0.939. 

\subsubsection{Jets of Radio Galaxies}

Radio galaxies form the other class of radio-loud AGN detected at TeV energies. Only four sources in this group have been detected in VHE gamma rays. Of these, three have been detected by VERITAS: M87, NGC 1275 and 3C 264, all belonging to the FR-I group of galaxies (after the Fanaroff \& Riley (1974) classification~\cite{Fanaroff_Riley}). In the unified model of AGN, it has been suggested that core-dominated FR-I and lobe-dominated FR-II radio galaxies form the parent populations of BL Lac objects and FSRQs, respectively, with the jets viewed at larger angles in the latter case~\cite{urry}. 3C 264 is the most distant radio galaxy detected at VHE, at a redshift of $z=0.0217$ (comoving distance of 93 Mpc). The discovery of TeV emission from this source by VERITAS during a flare in 2018, that lasted several weeks, triggered an extensive multiwavelength campaign, including VLBA, VLA, HST, Chandra and Swift~\cite{3c264}. The VERITAS and multiwavelength data showed that the flare detected in 2018 was associated with the unresolved core, with no detection of any enhanced activity in the  high-resolution Chandra or HST images compared with previous epochs. 

\subsubsection{Understanding Gamma-Ray Emission in Blazars} 

The measured spectral energy distribution (SED) of blazars and radio galaxies, plotted as $\nu F_\nu$ vs $\nu$ (where $F_\nu$ is the flux measured at the frequency $\nu$), or equivalently, $ E^2 dN/dE$ vs the energy $E$, is characterized by a double-peaked structure, where the lower-energy peak is due to synchrotron radiation of high-energy electrons in the  relativistically outflowing jet seen in blazars. 
In leptonic models, the higher-energy peak in the GeV-TeV band is generally attributed to inverse-Compton (IC) scattering of lower-energy photons off the population of relativistic electrons in the jet. Alternative hadronic models suggest that gamma-ray emission could result from energetic protons producing neutral pions (which decay to a pair of photons) in collisions with ambient gas. The protons could also upscatter low-energy photons in inverse-Compton processes.
Blazar spectra in the GeV-TeV band, as measured by VERITAS and other IACTs, help to differentiate between leptonic and hadronic scenarios. 

\begin{figure} [t!]
\centering
\vskip -0.1in
   \includegraphics[height=0.44\textwidth]{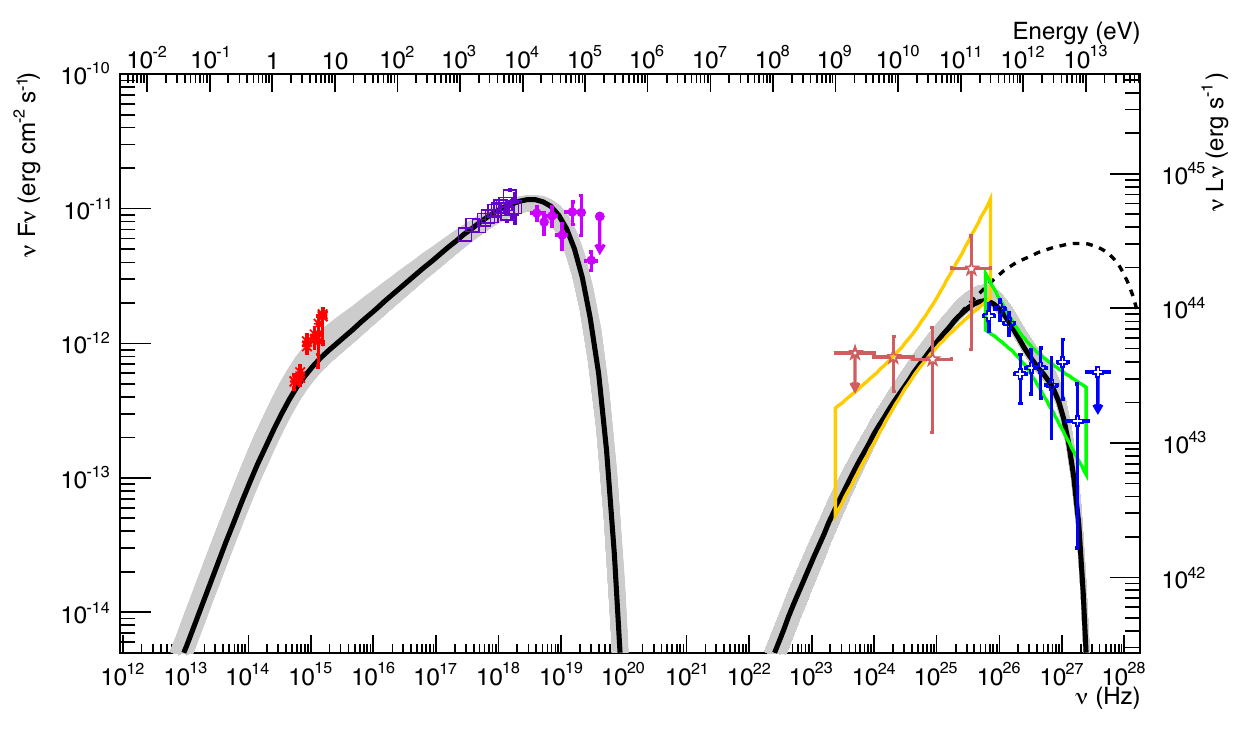}
  %\vskip -0.1in
     \includegraphics[height=0.42\textwidth]{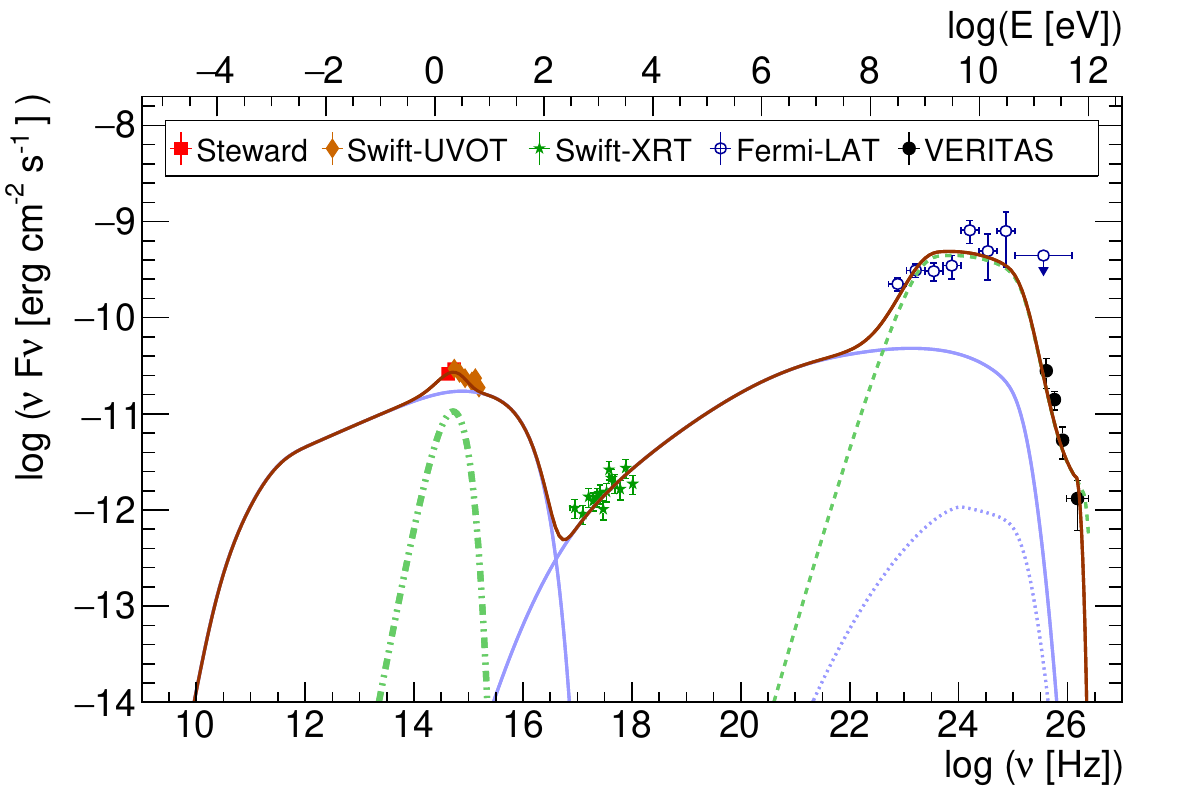}\vskip -0.1in
  \caption{ \small   
  (a) Upper: The multiwavelength SED of the HBL blazar 1ES 0229+200 from~\cite{1es0229_veritas}. The data are as follows: Swift-UVOT (red asterisks), Swift-XRT (open purple squares), Swift-BAT (pink circles), Fermi-LAT (salmon stars), and VERITAS (blue crosses). Note that the synchrotron peak is at X-ray energies, as is typically seen in HBLs. A one-zone SSC model shown in grey represents the data well. Further details may be found in the original reference. (b) Lower: The broadband SED of the FSRQ Ton~599 during the time of the VERITAS detection of the source in 2017 December. Overlaid (red curve)
  is a leptonic model calculation with an external radiation component. The various components of the model are as follows: synchrotron and SSC emission (solid blue curve), second-order self-Compton emission (dotted blue curve), thermal emission from the accretion disk (heavy dashed green curve), and inverse-Compton emission from the BLR (dashed green curve). Further details may be found in the original reference~\cite{ari_fsrq_2021}.}
  \label{fig:fig_SSC}
  \end{figure}

The HBLs detected by VERITAS largely show 
the synchrotron peak in the X-ray band and the higher-energy peak at TeV energies, suggesting that the synchrotron emission from 
ultrarelativistic electrons is  upscattered by the same population of electrons to high-energy gamma rays. Models most commonly used for explaining the emission in HBLs are the so-called ``SSC,'' or one-zone synchrotron self-Compton models (e.g.~\cite{maraschi1992}). Figure~\ref{fig:fig_SSC}a shows an example of a multiwavelength SED for the HBL 1ES 0229+200, where the one-zone SSC model of Katarzy\'nski et al.~\cite{katarzynski} was used to describe the spectrum~\cite{1es0229_veritas}. 1ES 0229+200 is a relatively distant blazar ($z = 0.1396$) and has a  hard gamma-ray spectrum (photon index $\Gamma \sim  2.5$). The figure demonstrates well-measured peaks in the X-ray and TeV bands, with the characteristic synchrotron peak at X-ray energies,  as typically seen in HBLs. 

In contrast to HBLs, Fig.~\ref{fig:fig_SSC}b shows an example of an SED for an FSRQ detected by VERITAS, Ton 599. FSRQs host radiatively efficient, or optically thick, accretion disks with an environment rich in UV-to-optical photons. External to the jet, FSRQs are believed to have ``clouds'' of gas in the so-called broad-line region (BLR) that can be photoionized by the  luminous UV continuum emitted by the disk, and a  ``dusty torus'' that is the source of IR radiation. TeV gamma rays are produced further out in the jet since  gamma-gamma interactions with low-energy photons make the base of the jet opaque to VHE photons.
The figure shows a physical model calculation to explain the SED of Ton 599 using a multicomponent SSC ``blob-in-jet'' model that includes an external IC-emission component from thermal accretion-disk radiation reprocessed by the BLR~\cite{hervet2015}. 

\subsubsection{Variability of Gamma-Ray Flux in Blazars}

\begin{figure} [t!]
\centering
%\vskip -0.1in
   \includegraphics[height=0.50\textwidth]{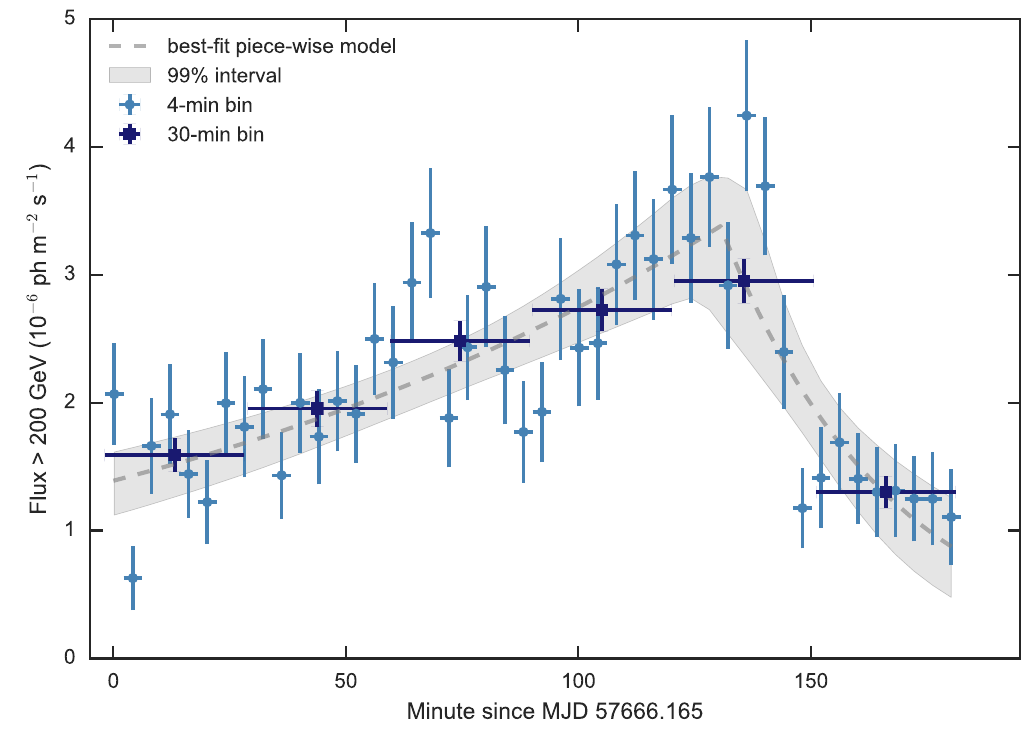}
    \includegraphics[height=0.30\textwidth]{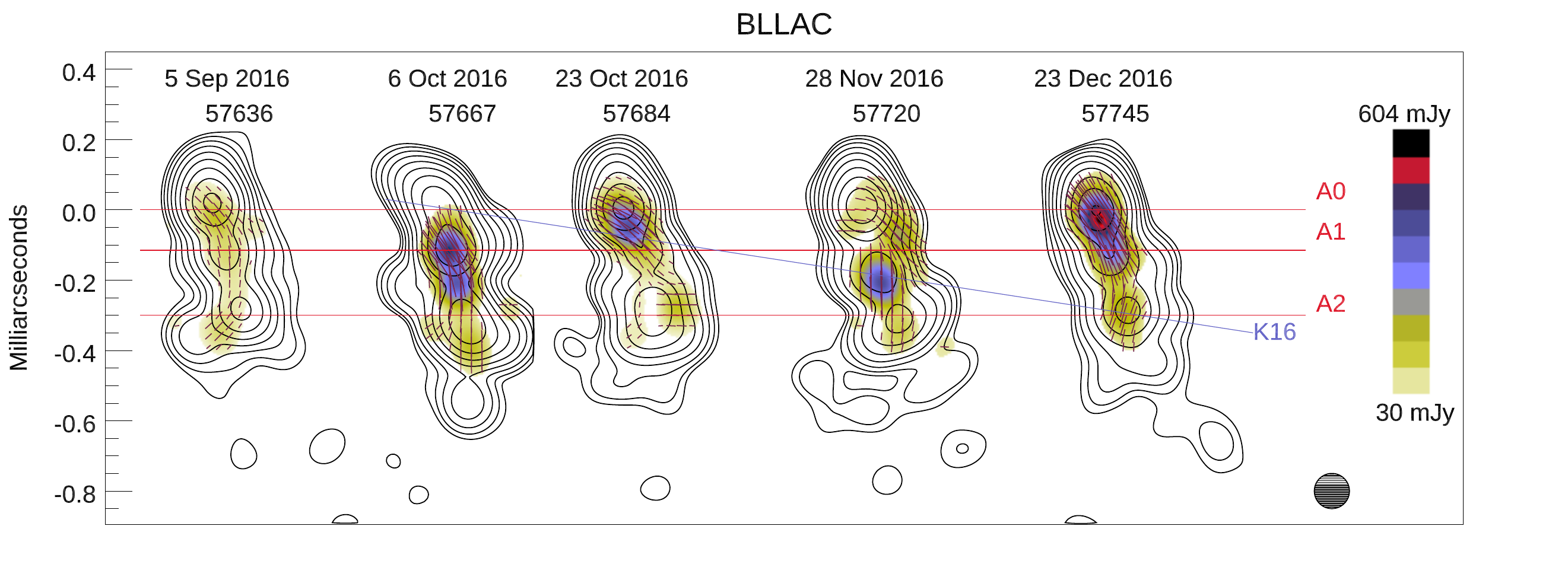}\vskip -0.1in
  \caption{ \small 
  (a) Upper: TeV light curves of BL Lac above 200 GeV measured by VERITAS during the flare in 2016 October. Data are shown in bins of 4-min (light-blue filled circles) and 30-min (dark-blue squares) bins. The grey dashed curve shows a model with the best-fit parameters and the shaded region illustrates the 99\% confidence interval, both of which are derived from simulations using Markov chain Monte Carlo sampling. (b) Lower: 43 GHz VLBA images of BL Lac showing the motion of the superluminal knot K16 across the epochs from 2016 October to December (blue curve). The total (contours) and polarized (color scale) intensity images of BL Lac are shown. (Details in~\cite{Abeysekara_2018_BLLAC}.) }
  \label{fig:bllac_flare}
  \end{figure}

Blazars are characterized by episodic flux variability across the electromagnetic spectrum. In the VHE band, VERITAS has detected variability on timescales of minutes to months in the gamma-ray flux. The physical origin of flux variability in blazars is not completely understood, but short-timescale variability places tight constraints on the size and location of the gamma-ray-emitting region~\cite{begelman}. 
Figure~\ref{fig:bllac_flare}a shows an example of a fast flare detected by VERITAS in the eponymous BL Lac object BL Lacertae in 2016~\cite{Abeysekara_2018_BLLAC}. A rapid rise time of about $2.3$ h in the gamma-ray flux was detected with a subsequent decay time of roughly $36$ min. The
peak flux above 200 GeV was measured to be approximately 1.8 times the flux level observed from the Crab Nebula. 

Figure~\ref{fig:bllac_flare}b shows results from coordinated VLBA observations of BL Lac at 43 GHz at epochs bracketing the time of the VERITAS flare, linking the ejection of a superluminal radio knot that passed the standing shock of the radio core of the jet around the time of the gamma-ray flare. This potentially provided seed photons for inverse-Compton scattering to gamma-ray energies. 
VERITAS and multiwavelength data were used to put %constraints 
limits on the Doppler factor of $\delta \geq 13$ and on the bulk Lorentz factor of $\Gamma \geq 7$, 
assuming a one-zone SSC model, the observed best-fit decay time of the TeV gamma-ray flare of 36 minutes, and a viewing angle of $2.2^\circ$~\cite{Abeysekara_2018_BLLAC}.

Blazar flares are unpredictable and occur on a wide range of timescales, offering clues to the physical processes in these objects. For example, data from the blazar OJ 287 were used to search for
evidence of periodic or quasiperiodic processes in blazars~\cite{OBrien2017}. In 
February 2014, VERITAS measured an isolated gamma-ray flare from the HBL B2 1215+30, that exhibited extreme luminosity with the TeV flux reaching about 2.4 times the Crab Nebula flux. This corresponded to an isotropic luminosity $\rm L_\gamma = 1.7 \times 10^{46}$ erg s$^{-1}$,
above an energy of 0.2 TeV, which was one of the highest to be ever observed from a TeV blazar. The variability timescale was measured to be less than 3.6 h, implying a Doppler factor of $\delta > 19$~\cite{1215_isolated}. 
Fast variability in blazars may be explained using models of magnetic reconnection events in the jets~\cite{Sironi_Spitkovsky2009}.  

\begin{figure} [t!]
\centering
\vskip -0.1in
\includegraphics[height=0.42\textwidth]
 {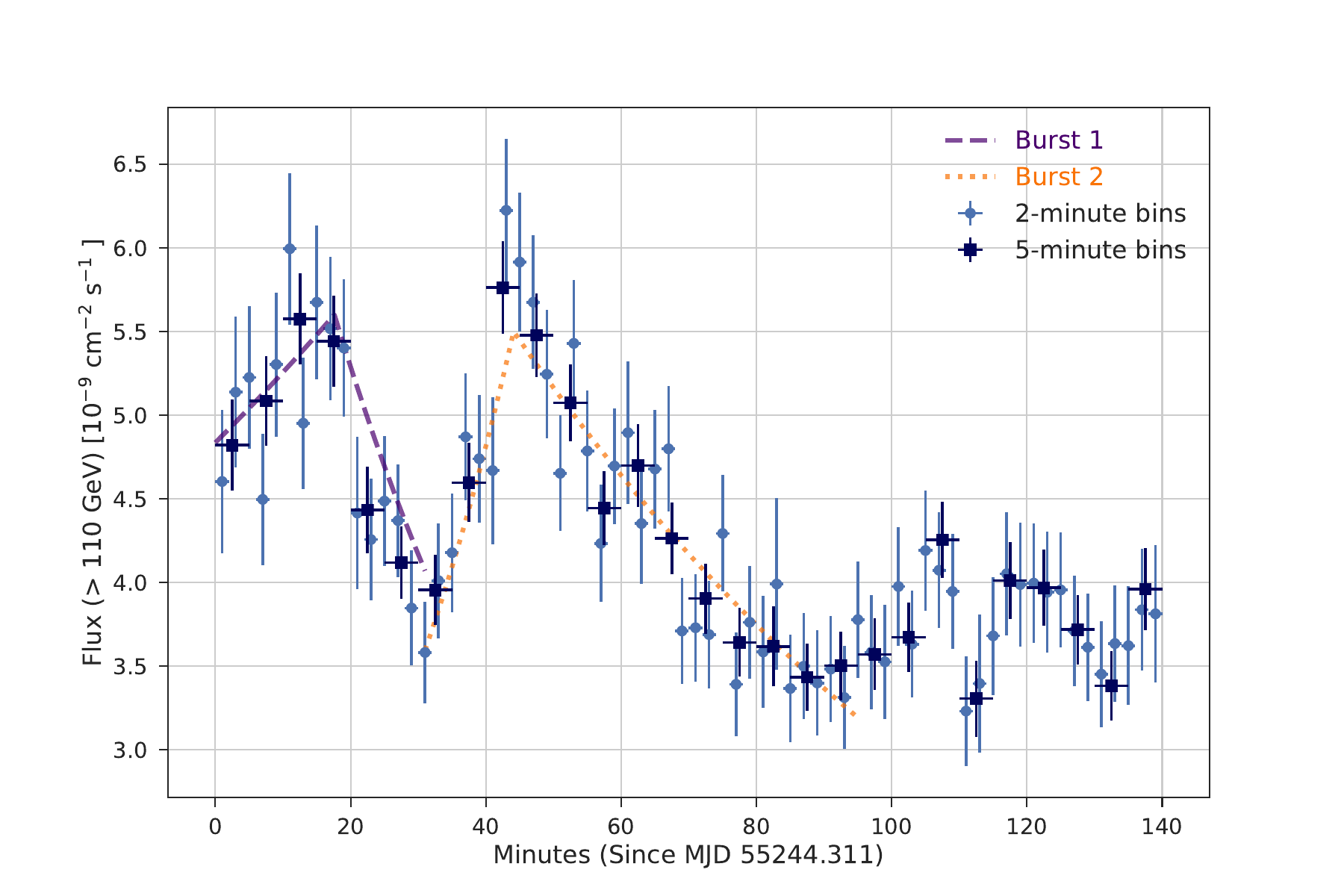}\vskip -0.05in
 \includegraphics[height=0.49\textwidth]{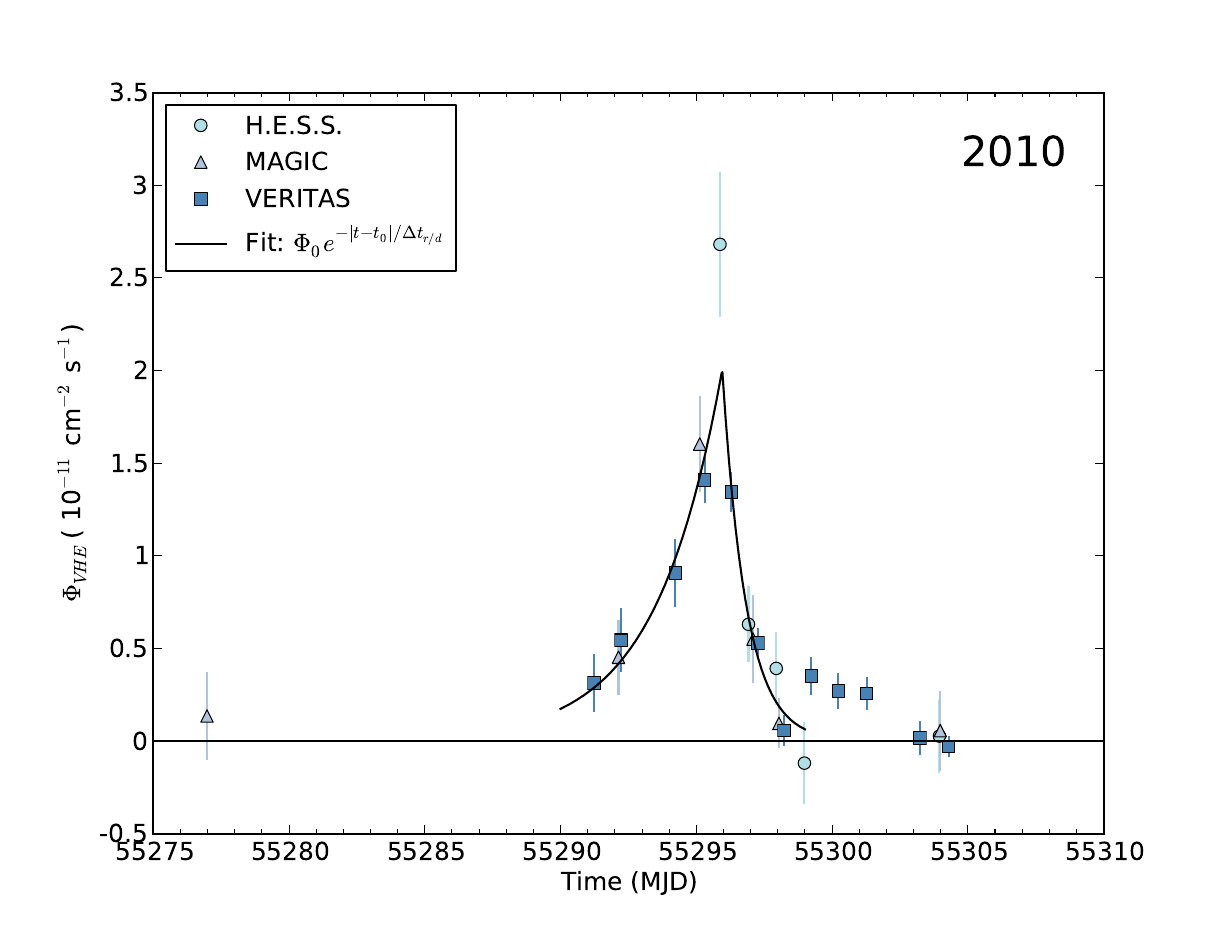}
\includegraphics[height=0.30\textwidth]{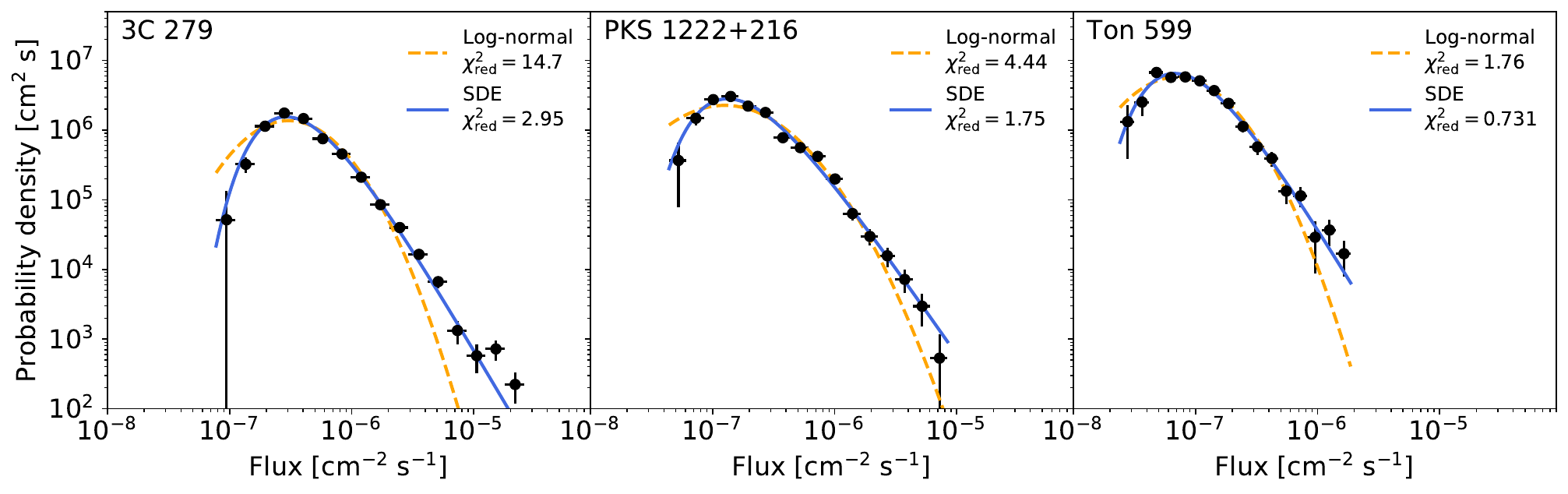}
  \caption{ \small  (a) Upper: The ``giant'' Mrk 421 flare detected by VERITAS in 2010, plotted using 2-min and 5-min bins, above an energy of 110
GeV. Two flares are identified using a Bayesian Block analysis. The dashed curves correspond to fits to an exponential function. (Figure from~\cite{mrk421_giantflare}). 
(b) Middle: A fast flare in the radio galaxy M87 detected simultaneously by VERITAS, H.E.S.S. and MAGIC in 2010. The fit is to a two-sided exponential function with short rise and fall times. (Details are in ~\cite{m87_2012_Abramowski}). 
(c) Lower: GeV-band flux distributions of the three FSRQs 3C 279, PKS 1222+216 and Ton 599, studied jointly with Fermi-LAT and VERITAS. Overlaid are fits with a log-normal probability density function (PDF)
(dashed orange) and a stationary-state PDF corresponding to a stochastic model (SDE)~\cite{tavecchio2020}, showing that in all cases, the stochastic model provides a better fit. (Figure from ~\cite{ari_fsrq_2021}.)}
\label{fig:mrk421_giant}
\end{figure}

A remarkable flare was measured by VERITAS in the historic TeV blazar Mrk 421~\cite{mrk421_giantflare} where it was possible to bin the data in 2-min timescales, as shown in Fig.~\ref{fig:mrk421_giant}a. This was one of the brightest flares ever observed in Mrk 421, with the peak gamma-ray flux 27 times the Crab Nebula flux, suggesting a Doppler factor $> 30$ for the blazar jet. Figure~\ref{fig:mrk421_giant}b shows another example of a well-sampled, single, isolated flare, with day-scale variability,  detected in 2010 in the radio galaxy M87 when coordinated observations were carried out by VERITAS, MAGIC and H.E.S.S. along with X-ray (Chandra) and radio (43 GHz, VLBA). The flare was well characterized by a two-sided exponential function, with a fall time that was a factor two shorter than the rise time and a peak flux that was about a factor 10 above the quiescent flux level of the source~\cite{m87_2012_Abramowski}. This was the first time an asymmetric flare profile was significantly detected in the source, and the variability timescale measured was the shortest ever detected in M87. 

Are blazar flares stochastic in nature? Figure~\ref{fig:mrk421_giant}c shows the Fermi-LAT flux distributions of three FSRQs from a recent study carried out jointly with VERITAS and Fermi-LAT, where modeling studies indicate that the data are best described by stochastic processes in the accretion disk~\cite{tavecchio2020}. Further studies of bright blazar flares, such as the ones carried out by VERITAS, will help elucidate the process of flux variations in blazars.

\subsubsection{Blazars as Probes of Cosmology} 

Gamma-ray observations of blazars have cosmological implications and the spectra of blazars may shed light on the density of EBL as well as on the nature of the intergalactic magnetic field (IGMF).
Gamma rays are probes for investigating Lorentz-invariance violation (LIV), exploring the existence of axion-like particles (ALPs) and 
studying cascades from ultra-high-energy cosmic rays (UHECR). The large VERITAS data set on TeV blazars is now being used to explore the details of their spectra~\cite{feng2021_tevspectra}. The study of intrinsic spectral curvature and spectral variability in blazars can help in understanding the physical mechanisms and the environment of TeV blazars,  and perhaps offer insights into exotic physics that may influence the opacity of the Universe to gamma rays.

\begin{figure} [t!]
\centering
%\vskip -0.1in
   \includegraphics[height=0.50\textwidth]{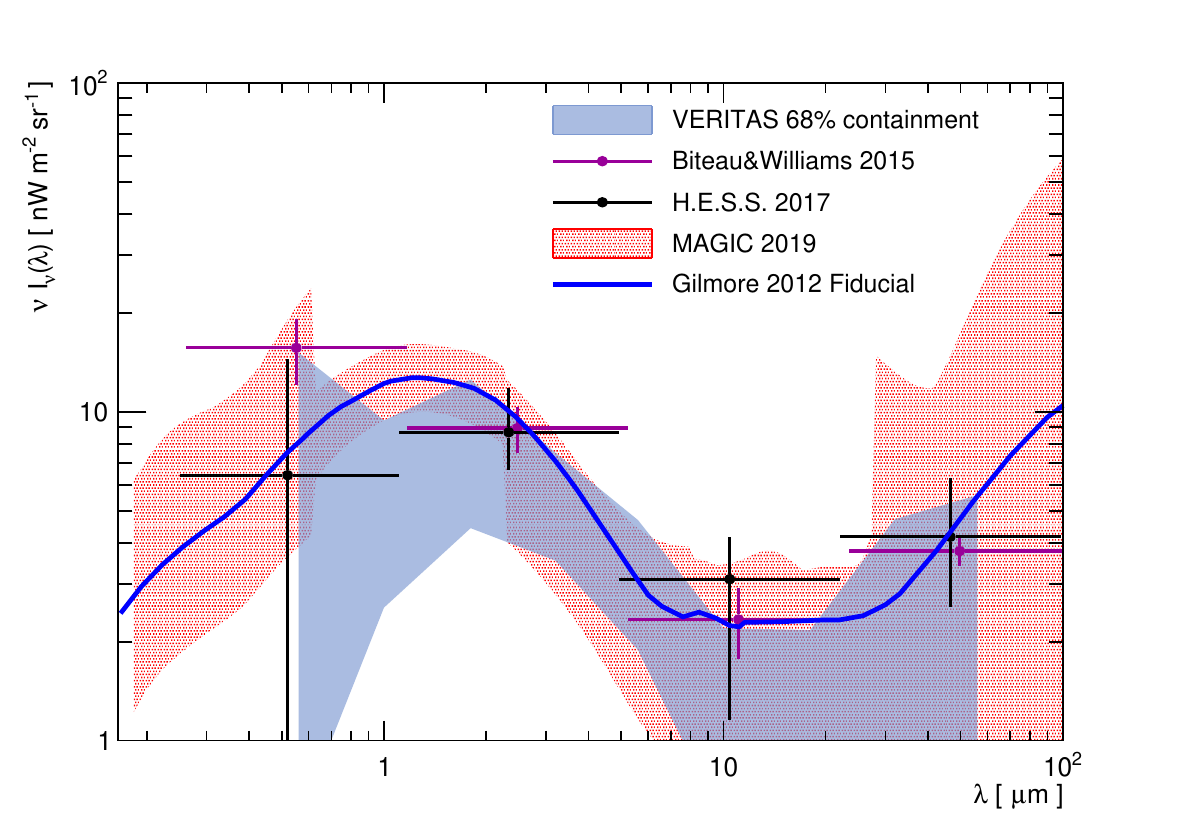}\vskip -0.1in
  \caption{ \small   
Plot showing the EBL intensity (68\% containment) measured by VERITAS as a function of wavelength, derived from nine years of VERITAS observations of 14 hard-spectrum blazars. The blazars are located at a redshift range of $z=0.044$ to $0.604$. Comparisons to other gamma-ray measurements are also shown. (Figure from~\cite{veritas_EBL_2019}.)}
  \label{fig:veritas_EBL}
  \end{figure}

Figure~\ref{fig:veritas_EBL} shows the spectral energy distribution of the EBL in the wavelength range of $0.56–56\ \mu$~m derived from VERITAS measurements of spectra of 14 hard-spectrum blazars observed over a period of nine years. The sources are located at redshifts ranging from $z=0.044$ to $0.604$. The results are found to be in good agreement with other recent measurements, 
and indicate that the EBL SED may be 
well described by lower limits assuming that the EBL is entirely due to radiation from cataloged galaxies~\cite{veritas_EBL_2019}. 

In another study, data on seven hard-spectrum blazars were analyzed to search for magnetically broadened gamma-ray emission from the sources, which may be seen in blazars due to cascade emission generated by the interaction of VHE photons with the EBL. In the presence of a non-zero intergalactic magnetic field (IGMF), the electron-positron pairs are deflected producing a detectable angular broadening in the emission seen in the blazars. The VERITAS study carried out dedicated Monte Carlo simulations of the angular profiles of the sources to compare with observations. Based on the analysis, an IGMF strength around $10^{-14}$ G was excluded at the 95\% confidence level~\cite{veritas_igmf_2017}. 

\subsubsection{The Starburst Galaxy M82}

M82 is the fist starburst galaxy to be associated with VHE gamma-ray emission. VERITAS detected the source in observations carried out in 2008 and 2009 and measured a flux of approximately 0.9\% of the Crab Nebula flux, making it one of the weakest sources to be detected by VERITAS~\cite{m82nature2009}. Starburst galaxies are attractive targets for studying the origin of cosmic rays due to their exceptionally high rates of star formation, and potentially high cosmic-ray density which can lead to diffuse gamma-ray production via cosmic ray interactions with interstellar gas and radiation. The VERITAS measurements indicated a cosmic-ray density of 250 eV cm$^{-3}$ in the starburst core of M 82, much higher than the average density in the Milky Way Galaxy.

\subsection{Galactic Astrophysics}

VERITAS, being located in the northern hemisphere, has a limited view of the Milky Way Galaxy. The majority of Galactic targets are visible from the VERITAS observatory only in late spring and early summer months. During the summer, observation times are typically short, and additionally, the VERITAS observatory shuts down during the summer monsoon season in July and August. The source count of Galactic targets detected by VERITAS is smaller than that in the extragalactic catalog, but nevertheless, VERITAS has a robust Galactic observing program with impactful results in key areas. Galactic science with VERITAS covers the following broad areas: 

\begin{itemize}
    \item 
the study of Galactic binaries, 
\item the search for Galactic cosmic rays and the study of supernova remnants (SNRs) including nonthermal shells and shell-molecular cloud interactions, 
\item the observation of TeV pulsar wind nebulae (PWNe) and the search for PeVatrons, including the study of unidentified sources detected by wide-field arrays (such as HAWC or LHAASO), 
\item surveys of the Galactic plane and the search for diffuse emission, and 
\item the study of pulsed emission from the Crab pulsar and the search for pulsed emission from other pulsars.
\end{itemize}

A particular highlight from VERITAS was the discovery of pulsed emission from the Crab pulsar, for the first time, at energies above 100 GeV. 
The following subsections summarize a few of the key results from VERITAS. 

\subsubsection{Supernova Remnants}

SNRs have been suggested as natural sites in which strong shocks develop with high speeds and where cosmic ray acceleration can take place. How these sources accelerate particles and whether particles are accelerated to petaelectronvolt (PeV) energies are still open questions. VERITAS has acquired deep observations of three northern SNRs: Tycho, Cas A and IC 443. 

\begin{figure} [t!]
\centering
\vskip -0.1in
\includegraphics[height=0.5\textwidth]
 {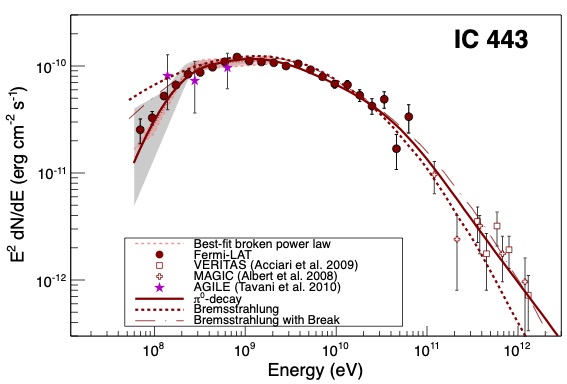}
\includegraphics[height=0.55\textwidth]{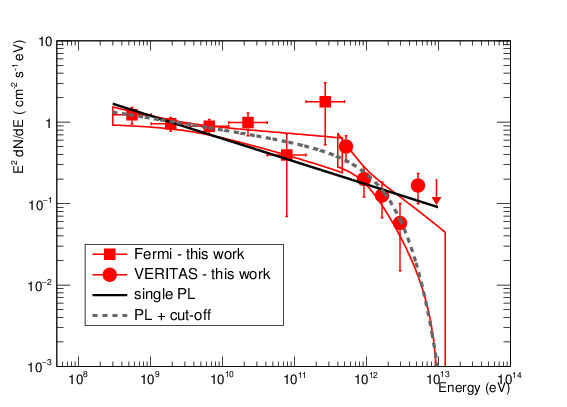}\vskip -0.1in
\caption{\small  (a) Upper: Spectrum of the SNR IC 443 measured at gamma-ray energies by Fermi-LAT, VERITAS and MAGIC. The curves and shaded areas represent fits to the Fermi-LAT spectral data only. The solid curve is the best fit to a pion-decay model for gamma-ray production. The TeV data points are not taken into account for the fit. (Figure is from~\cite{fermi_ic443_2013} where further details may be found). %Fig 2A https://arxiv.org/abs/1302.3307 
(b) Lower: Gamma-ray spectrum of the SNR Tycho measured by Fermi-LAT and VERITAS showing fits both to a single power-law model and to a power-law model with a cutoff (dashed curve). (Figure and details of the fit are  in~\cite{veritas_tycho_2017}.)}
\label{fig:ic443_tycho_sed}
% (b) Fig. 6 in https://arxiv.org/pdf/1701.06740.pdf
\end{figure}

Figure~\ref{fig:ic443_tycho_sed}a shows the gamma-ray spectrum of the middle-aged SNR IC 443, as measured with Fermi-LAT, overlaid with data from MAGIC and VERITAS.
The Fermi-LAT spectrum is well fitted by a model that assumes IC 443 is a hadronic accelerator~\cite{fermi_ic443_2013}. The gamma rays are produced by the decay of neutral pions produced in hadronic interactions with target material in giant molecular clouds in the vicinity. These data provide 
evidence that protons are accelerated in SNRs, which has long been a key issue in understanding the origin of cosmic rays. VERITAS data, however, show that at the highest energies, the spectrum cuts off above $100$ GeV, making it problematic to explain the entire Galactic cosmic ray population up to the “knee” at $3 \times 10^{15}$~eV. 

The softening of the spectrum at very high energies is also seen in the case of Tycho's SNR (also known as Tycho, SNR G120.1+1.4), a young, type Ia SNR, located in a relatively clean environment, that has been well-studied across all wave bands. Initial measurements by VERITAS indicated that the spatial distribution of the gamma-ray emission can be explained as either a point source or consistent with a uniform SNR shell origin~\cite{veritas_tycho_2011}. Figure~\ref{fig:ic443_tycho_sed}b shows the gamma-ray spectrum of Tycho obtained from about 150 hours of data, along with measurements by Fermi-LAT~\cite{veritas_tycho_2017}. A steepening of the spectrum is detected at energies above $400$ GeV, putting a constraint on the maximum energy of accelerated particles, and suggesting again that it is unlikely that SNRs power the whole Galactic cosmic-ray population. Several earlier models for the spectral energy distribution seen in Tycho suggest that the gamma-ray emission is largely due to hadronic interactions (e.g.~\cite{morlino_caprioli2012,slane2014}), with maximum proton energies greater than 50 TeV. The spectral softening detected by VERITAS, however, indicates that the maximum particle energies in Tycho are lower than that predicted by theoretical models (see discussion in~\cite{veritas_tycho_2017}. 

VERITAS has extensively observed the SNR Cas A, a source that was one of the early VHE sources, first detected by the HEGRA stereoscopic Cherenkov telescope system~\cite{hegra_casa_2001}. Cas A is a young SNR, remnant of a Type IIb supernova explosion, that has been studied across the electromagnetic spectrum from radio to gamma rays. The size of the remnant as measured at X-ray and radio energies is comparable to the point-spread function (PSF) of current IACTs, and Cas A is unresolved so far at gamma-ray energies.
VERITAS has observed the source at various times since 2007, covering the energy range from 200 GeV to 10 TeV. The data in the Fermi-LAT range show significant spectral curvature centered on an energy of $1.3\pm 0.4_{\rm stat}$ GeV, evidence of a pion bump from hadronic processes. The gamma-ray spectrum measured by Fermi-LAT and VERITAS shows that the joint spectrum deviates from a simple power-law function. A cutoff is seen at energies above $2.3 \pm 0.5_{\rm stat}$ TeV~\cite{veritas_casa_2020}.  

\begin{figure} [t!]
\centering
\includegraphics[height=0.6\textwidth]{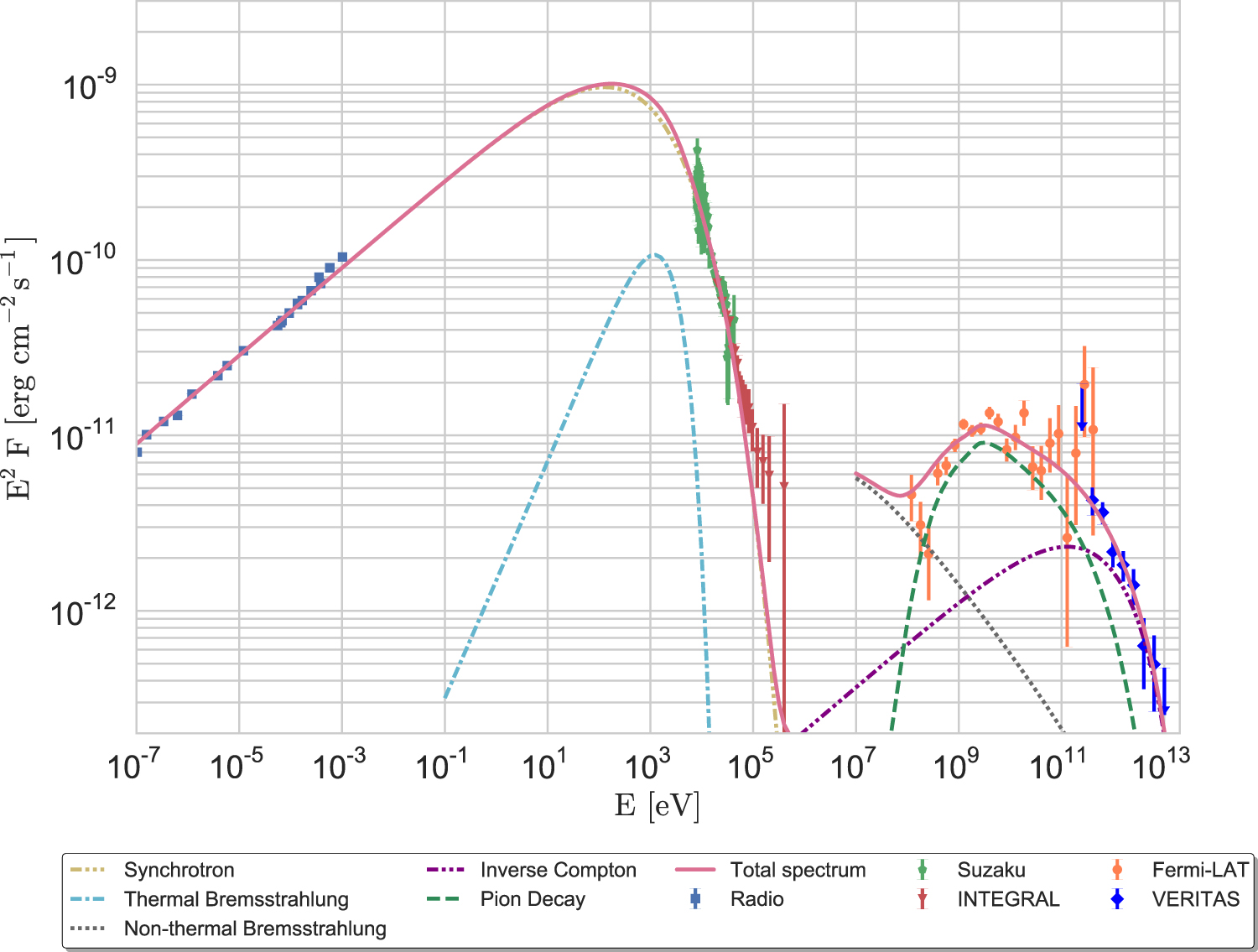}\vskip -0.1in
\caption{\small Broadband spectral energy distribution of the SNR Cas A showing the data taken with VERITAS and at other energies. The curves correspond to a leptohadronic model which best describes the data. 
(Figure from ~\cite{veritas_casa_2020} where further details may be found.)
}
\label{fig:casa_sed}
% Fig. 9 in https://arxiv.org/pdf/2003.13615.pdf
\end{figure}

Together with IC 443 and Tycho, Cas A does not qualify as a ``PeVatron,'' a cosmic accelerator capable of accelerating protons to peV energies, 
although the gamma-ray spectrum of Cas A suggests that it is a hadronic accelerator.  Figure~\ref{fig:casa_sed} shows a spectral energy distribution for Cas A from radio to gamma-ray wavebands, which was analyzed using several different emission models. The VERITAS observations do not indicate any evidence for a nonthermal bremsstrahlung flux above 100 MeV. Based on spectral modeling, a purely one-zone leptonic scenario was ruled out, suggesting 
 proton acceleration up to TeV energies. The figure shows a leptohadronic model explaining the observed gamma-ray spectrum, in which protons are accelerated up to at least 6 TeV. (See details in ~\cite{veritas_casa_2020}).

In the study of SNRs, a key result from VERITAS was the detection of extended emission from IC 443, which VERITAS was able to resolve on a scale of a few arcminutes. IC 443 is the only northern SNR to be clearly resolved in VHE gamma rays. 
The VERITAS measurements of the extent match the Fermi-LAT measurements, with the morphology coinciding with shocked-gas distribution, indicating a uniform cosmic ray acceleration throughout the shell~\cite{humensky_ic443_2015}. 

\subsubsection{Pulsar Wind Nebulae and the Search for PeVatrons}

The search for PeV ($10^{15}$ eV) emission from astrophysical sources has been one of the highest priorities for ground-based gamma-ray experiments (see~\cite{cristofari_2021} for a review). The hunt for PeVatrons was recently successfully carried out by the LHAASO Collaboration who reported the detection of a dozen sources with spectra extending beyond 100 TeV~\cite{lhaaso_pev_2021}, as well as by HAWC that reported the detection of several PeVatrons~\cite{hawc_pev_2020}.  
Supernova remnants and pulsar wind nebulae (PWNe), with relativistic plasma, 
are the best candidates to target in a search for
PeVatrons. The Crab Nebula, the prototypical PWN, is a confirmed PeVatron, as reported by LHAASO~\cite{lhaaso_crab_2021}.  
Somewhat surprisingly, the three supernova
remnants detected by VERITAS show spectral cutoffs at high energies and are therefore unlikely to be the source of $10^{15}$ eV cosmic rays; i.e. PeVatrons (see previous discussion). 
In VERITAS data, two promising candidates for PeVatrons are VER J2227+608/SNR G106.3+2.7 and MGRO J1908+06, and a deep study of these two sources is currently underway. 

\begin{figure} [t!]
\centering
%\vskip -0.1in
\includegraphics[height=0.5\textwidth]
 {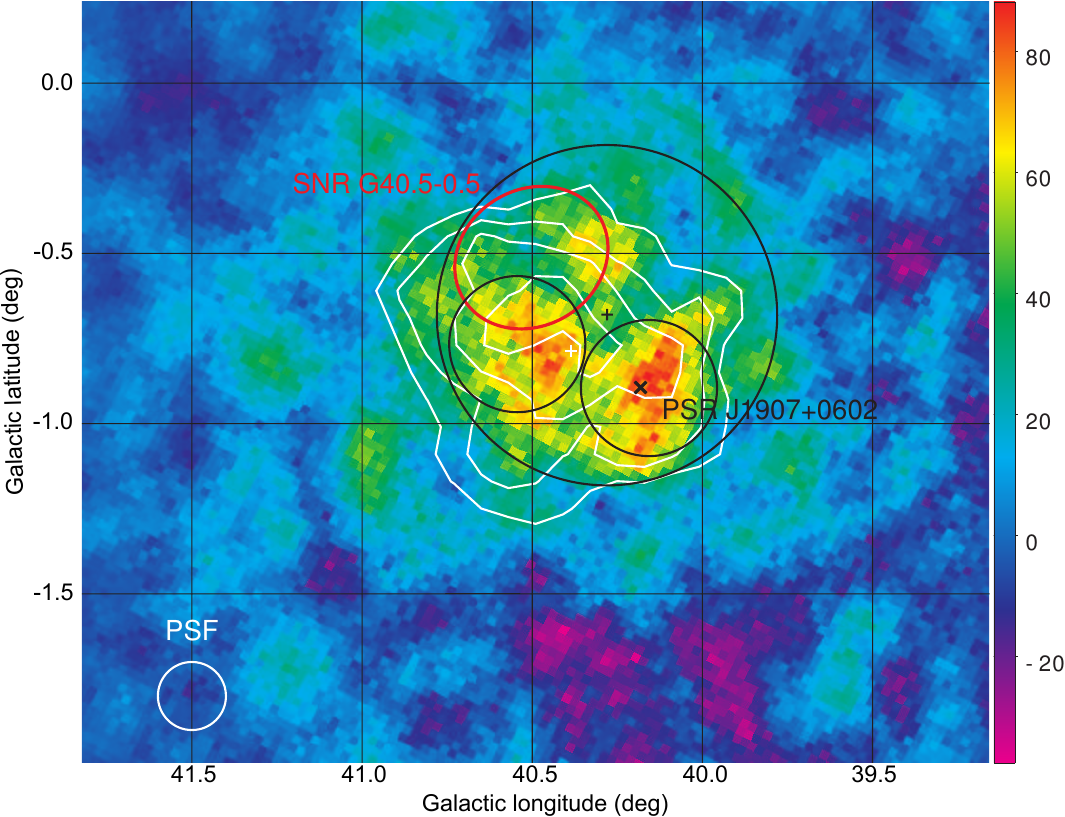}
 \includegraphics[height=0.52\textwidth]
 {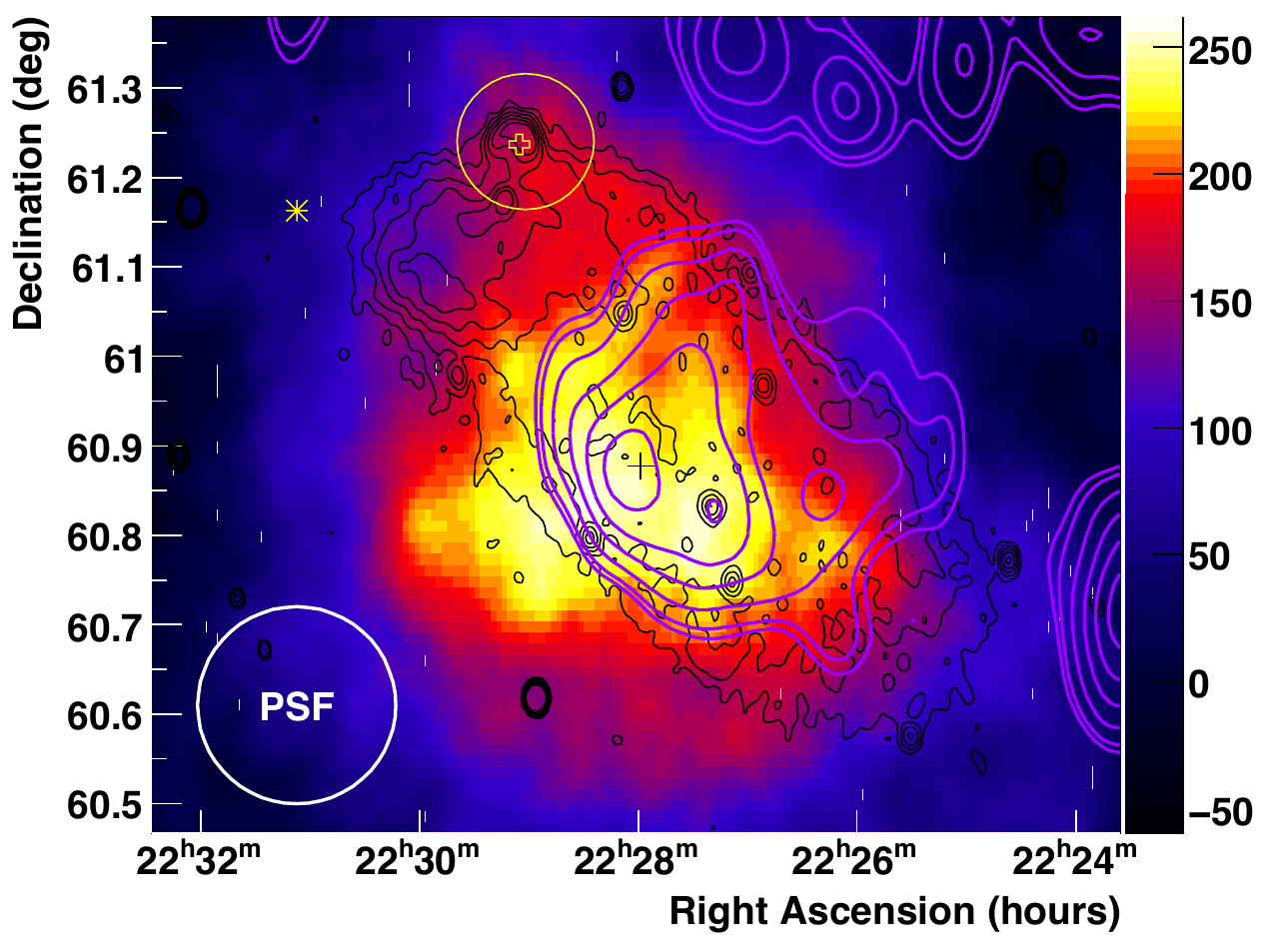}%https://arxiv.org/abs/0911.4695 Fig. 1
 \vskip -0.1in
  \caption{ \small  
 (a) Upper: sky map of the complex region around MGRO J1908+06 showing the excess gamma-ray counts detected by VERITAS. The black
cross represents the best-fit location obtained from the VERITAS data. 
The red ellipse shows the extent of SNR G40.5–0.5 in radio continuum emission and the 
black ‘x’ shows the location of PSR J1907+0602. The overlaid white contours are the 4 to 7 $\sigma$ significance levels from the excess map measured by H.E.S.S. with the white cross showing the H.E.S.S. best-fit position. Details and discussions are in ~\cite{veritas_mgroj1908_2014}. (b) Lower: Sky map of excess gamma-ray events emission from G106.3+2.7, as measured by VERITAS. The black cross marks the centroid of the TeV emission. 
The location of pulsar PSR J2229+6114 is marked by the open yellow cross. The boomerang-shaped wind nebula is evident in black contours showing the 1420 MHz radio emission. 
(Figure from~\cite{veritas_boomerang_2009}, where further details may be found.)}
\label{fig:mgroj1908_boomerang}
\end{figure}

Photons greater than $100$ TeV have been detected from  MGRO J1908+06 by both HAWC and LHAASO. MGRO J1908+06, associated with the SNR G40.5-0.5, is a powerful extended gamma-ray source discovered by Milagro~\cite{milagro_1908_2007}. The VERITAS sky map, shown in Figure~\ref{fig:mgroj1908_boomerang}a, suggests a complex region with TeV emission due to energetic particles, emitted by the pulsar PSR J1907+0602, interacting with either the SNR or molecular clouds~\cite{veritas_mgroj1908_2014}. The TeV emission detected by VERITAS is strongest in the region
near PSR J1907+0602 and also extends towards the SNR G40.5–0.5.

VER J2227+608, associated with the tail region of the SNR G106.3+2.7, is another source identified as a PeVatron in the LHAASO data~\cite{lhaaso_pev_2021}. 
The broadband gamma-ray data at Fermi-LAT and TeV energies were recently reported to be well fitted by a single power-law function with an index of
$1.90\pm0.04$, without any evidence of spectral cut off~\cite{xin_boomerang_2019}. A modeling study found that the cutoff energy for the proton distribution needs to higher than $\sim 400$ TeV to explain the gamma-ray  spectrum~\cite{xin_boomerang_2019}. Similarly, HAWC reported measuring a hard spectrum in the 40 to 100 TeV band with no evidence for a cutoff~\cite{hawc_boomerang}. VERITAS previously measured VHE emission from the source as shown in Fig.~\ref{fig:mgroj1908_boomerang}b. The source was found to be extended, with the centroid of the VHE emission $0.4^{\circ}$ away
from the pulsar PSR J2229+6114, which is located at the northern edge of the SNR~\cite{veritas_boomerang_2009}. The associated radio and X-ray-emitting wind nebula is boomerang-shaped, hence the name ``Boomerang" is often used for the source. 

\subsubsection{The Cygnus Survey \& the Galactic Diffuse Emission}

\begin{figure} %[H]
\centering
%\vskip -0.1in
\includegraphics[height=0.28\textwidth]
{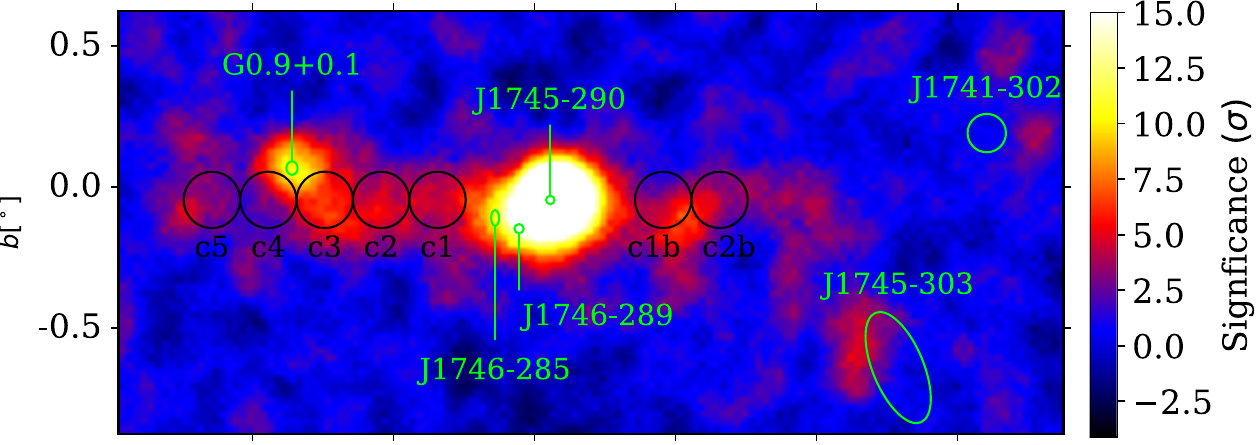}
\includegraphics[height=0.28\textwidth]
 {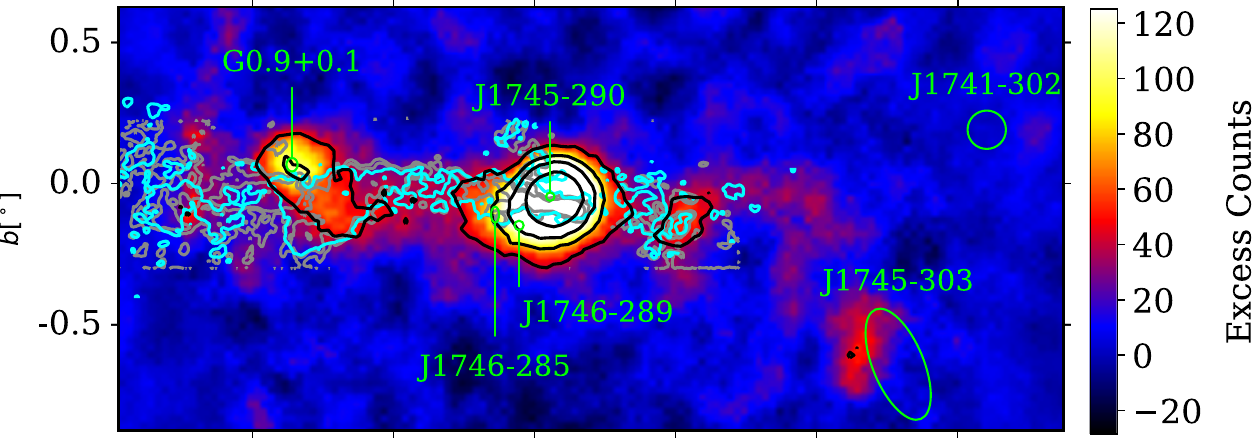}
\includegraphics[height=0.33\textwidth]
 {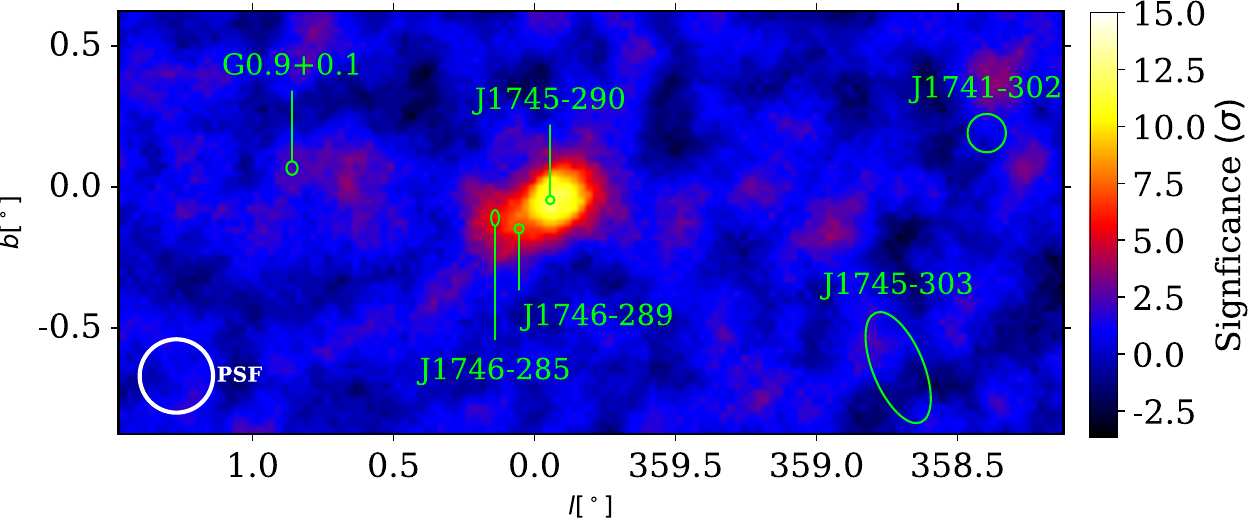}
 %Fig. 1 middle and bottom https://arxiv.org/pdf/2104.12735.pdf
 \vskip 0.1in
 \includegraphics[height=0.4\textwidth]
 {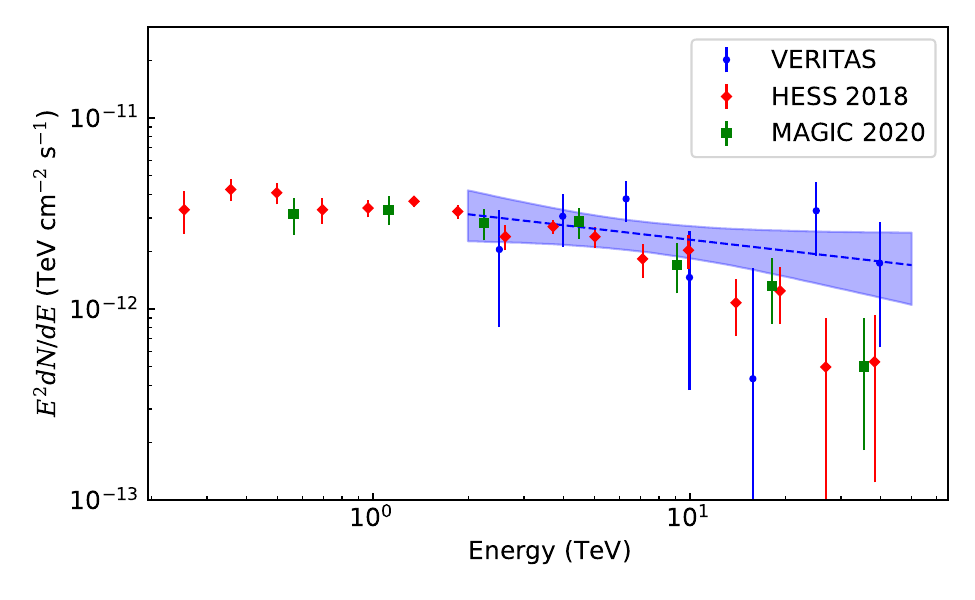}%Fig. 5 https://arxiv.org/pdf/2104.12735.pdf
 %\vskip -0.1in
  \caption{ \small  
(a) Upper: Statistical significance maps of gamma-ray-like events above 2 TeV (top panel) and 10 TeV (bottom panel), and map of excess gamma-ray counts above 2 TeV for the Galactic Center region and Galactic ridge (middle panel), as measured by VERITAS. Locations and 68\% confidence regions for known point sources are shown as green ellipses. Further details and image descriptions are in~\cite{veritas_GC_2021}. 
(b) Lower: Differential energy spectrum of the Galactic ridge diffuse emission measured by VERITAS (blue) overlaid with the spectra measured by H.E.S.S. (red) and MAGIC (green). Note that the VERITAS data points are scaled up to account for the larger regions modeled by H.E.S.S. and MAGIC for the diffuse ridge spectrum. The best-fit (from VERITAS data) power-law function is shown as the blue dashed curve together with the $1\sigma$ confidence
band on the model fit (shaded region).
(Figure from~\cite{veritas_GC_2021}, where further details may be found.)}
\label{fig:veritas_gc}
\end{figure}

VERITAS has detected diffuse gamma-ray emission from the Galactic plane, measuring a hard spectrum to 40 TeV with no evidence of a break. Figure~\ref{fig:veritas_gc}a shows the excess gamma-ray events above 2 TeV in the inner $2^{\circ} \times 1.25^{\circ}$ region of the Galactic Center measured by VERITAS using 125 hours of data taken between 2010 and 2018~\cite{veritas_GC_2021}. Strong diffuse emission is detected along the Galactic ridge, as shown in the figure. The central source in the figure is VER J1745–290, which is  consistent with the position of Sagittarius A* and is detected with a high statistical significance. Figure~\ref{fig:veritas_gc}b shows the spectrum of the diffuse emission along the Galactic ridge.  The spectrum is seen to be hard, and is best fitted by a power-law with an index of $2.19 \pm 0.20$ with no cutoff seen up to 40 TeV. In comparison, the spectrum of VER J1745–290 shows a softening at high energies, with an exponential cutoff at $\sim 10$ TeV~\cite{veritas_GC_2021}. The Galactic ridge results offer strong evidence for a potential PeVatron at the Galactic Center.

\begin{figure} [t!]
\centering
\includegraphics[height=0.5\textwidth]
 {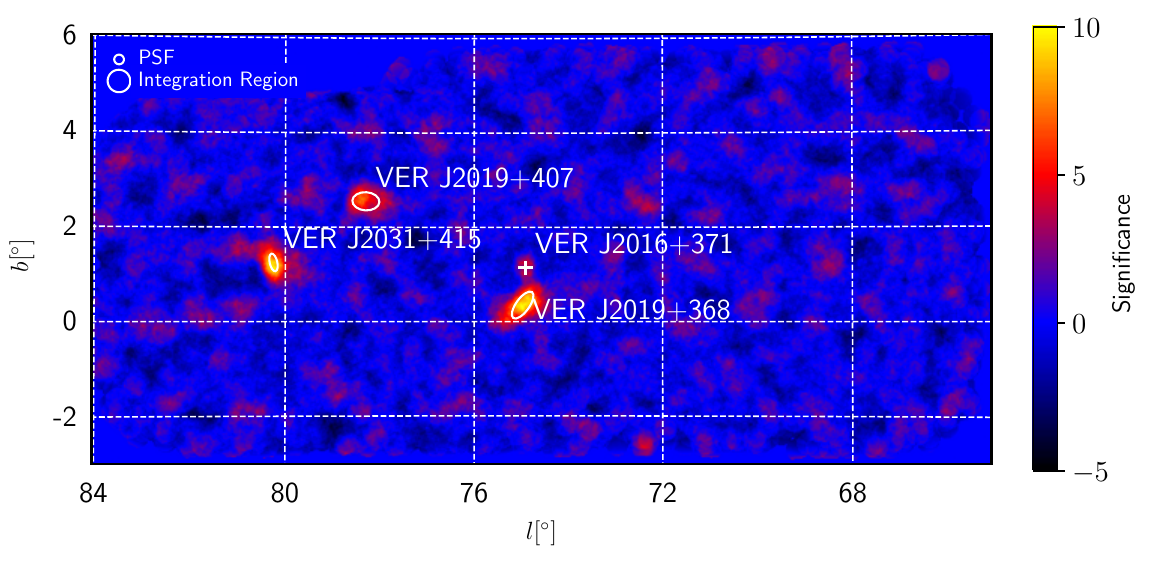}
 \vskip -0.1in
  \caption{ \small  
VERITAS sky map of the Cygnus region showing significance of gamma-ray detection above the background for energies above 400 GeV. The analysis used an integration region of  $0.23^{\circ}$ to search for extended sources. Three extended sources and one point source were detected in the survey. The overlaid white ellipses indicate the $1\sigma$ extent for the source sizes fitted with a two-dimensional Gaussian function. 
(Figure from~\cite{veritas_cygnus_abeysekara2018}, where further details may be found.)}
\label{fig:veritas_cygnus}
\end{figure}

A survey of the Cygnus region by VERITAS  
covered a $15^{\circ}$ by $5^{\circ}$ portion of the Galactic plane, from $67^{\circ}$ to $82^{\circ}$ in (Galactic) longitude and from $-1^{\circ}$ to $+4^{\circ}$ in latitude.
The motivation for the survey was to explore a region that is a perfect laboratory for the study of cosmic rays and represents the largest and most active region of creation and destruction of massive stars in the Milky Way. The Cygnus region is also the brightest region of diffuse gamma-ray emission in the northern sky. Figure~\ref{fig:veritas_cygnus} shows results from the VERITAS survey carried out over a period from 2007 to 2012, using more than 300 hours of data, which included targeted follow-up observations of some sources. The figure shows a complex region with three extended sources (VER J2019+407, VER J2031+415, VER J2019+368) and one point source (VER J2016+371) detected by VERITAS~\cite{veritas_cygnus_abeysekara2018}.

%%%%%%
\subsubsection{The Crab Pulsar}

\begin{figure} %[t!]
\centering
\vskip -0.1in
\includegraphics[height=0.55\textwidth]
% {Figures/Crabpulsar_SED_Aliu2011.png} %Fig.2
{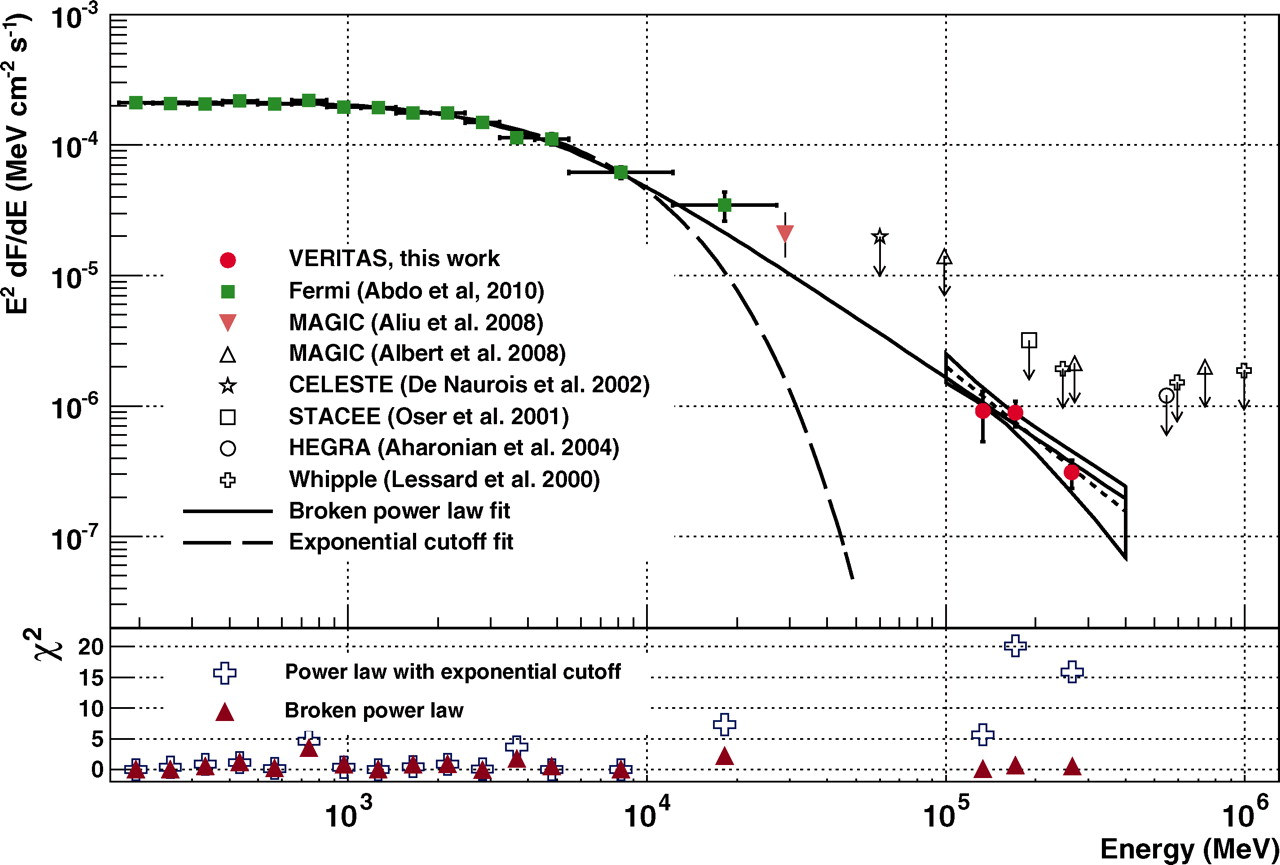} %Fig.2
\includegraphics[height=0.42\textwidth]{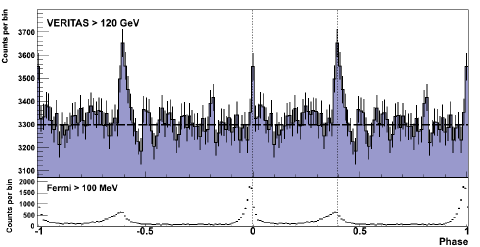} % Fig 1
\vskip -0.1in
  \caption{ \small  (a) Upper: The energy spectrum of the Crab pulsar measured from MeV to GeV energies using satellite and ground-based gamma-ray detectors. The VERITAS measurements between 100 and 400 GeV are shown, along with other historical measurements. (Figure from~\cite{veritas_crab}). (b) Lower: Pulse profile of the Crab pulsar as measured by VERITAS and Fermi-LAT. The peak positions of the main pulse (P1) and the interpulse (P2) are shown. The vertical dashed lines mark the best-fit peak position of P1 and P2 from the VERITAS data. Details are in the original reference~\cite{veritas_crab}.}
\label{fig:crab_pulsar}
\end{figure}

A key result from VERITAS was the detection of pulsed VHE emission from the Crab pulsar. Earlier observations by satellite experiments EGRET and Fermi-LAT showed that the Crab pulsar emitted pulsed gamma rays up to at least 25 GeV~\cite{egret_crab, fermi_crab}. 
The first detection with Cherenkov telescopes of pulsed gamma-ray emission above 25 GeV from the pulsar was reported by MAGIC, and suggested that the emission zone is far out in the pulsar magnetosphere~\cite{magic_crab}. In 2011 VERITAS reported the first detection of pulsed emission between 100 and 400 GeV~\cite{veritas_crab} from the source. VERITAS measured the spectrum of the Crab pulsar and showed that the spectrum in the 100-400 GeV range is best described by a broken power-law function. These data provide strong  constraints on the site and production mechanisms of the gamma rays, and models suggest that curvature radiation is unlikely to be primarily responsible for the detected gamma-ray emission~\cite{lyutikov2012}. Figure~\ref{fig:crab_pulsar}a shows the spectral energy distribution for the Crab pulsar, including the VERITAS measurements as well as archival studies at lower energies. Figure~\ref{fig:crab_pulsar}b shows the pulse profile for the Crab pulsar as measured by VERITAS (shaded region) above 120 GeV and by Fermi-LAT above 100 MeV. 
The two main features in the pulse
profile of the Crab pulsar as measured across the electromagnetic spectrum, the main pulse (P1) and the interpulse (P2), are seen in the figure. The VERITAS and Fermi-LAT measurements of the positions of these features agree within uncertainties. The VERITAS data show a peak amplitude at phase 0.4. The phase-folded pulse profile measured by VERITAS was found to be narrower by more than a factor of two than that seen at 100 MeV by Fermi-LAT~\cite{veritas_crab}.

The detection by VERITAS presents a challenge to current pulsar models. The results suggest that the gamma rays likely originate close to, or outside of, the light cylinder—beyond 10 stellar radii from the neutron star. It is possible that the VHE emission is due to inverse-Compton scattering, as discussed in~\cite{veritas_crab}. Since these findings, the Crab pulsar was subsequently seen to emit pulsed emission above 400 GeV, with the spectrum extending to beyond 1.5 TeV~\cite{ansoldi2016}. 

\subsubsection{Gamma-Ray Binaries} 

Binary star systems are the only point-like and variable TeV sources (other than pulsars) in the Galaxy, and just 11 have been detected at very high energies. VHE-gamma-ray binaries are thought to be systems comprising a massive O or Be star and a compact object such as a young pulsar. The gamma-ray emission is likely powered by the spin-down energy of the pulsar~\cite{dubus2013}. However, the nature of the compact object is not known for many of these systems. 

VERITAS has detected four gamma-ray binaries to date, namely, \lsi, HESS J0632+057, TeV J2032+4130, and LS 5039, each one of which has distinctive characteristics. VERITAS observations help address several open questions in the study of gamma-ray binaries: whether binaries are powered by pulsar-wind or accretion systems, what drives the variability seen in binaries, what is the role of stellar winds, and most of all, why are only a few high-mass X-ray binaries detected at gamma-ray energies?

VERITAS data were key to the identification of the unidentified TeV source, HESS J0632+057, first detected by H.E.S.S. in the Monoceros region, as a gamma-ray binary. Dedicated long-term Swift (X-ray) and VHE gamma-ray data over ten years helped in the measurement of the binary period. Initial H.E.S.S. data identified the source to be point-like~\cite{hess_0632_2007} and subsequent VERITAS observations provided evidence that the source is variable, suggesting that it could be an undiscovered gamma-ray binary~\cite{veritas_0632}. Follow-up long-term X-ray observations were carried out and measurement of timing and variability information in X-rays showed the orbital period of the binary to be $315 \pm 5$ days~\cite{bongiorno2011, aliu_0632_2014}, spectacularly confirming the binary hypothesis. The compact object was identified as MWC 148, a massive
emission-line star, of spectral type
B0pe~\cite{hinton2009}. 

\begin{figure} [b!]
\centering
\vskip -0.1in
\includegraphics[height=0.48\textwidth]
 {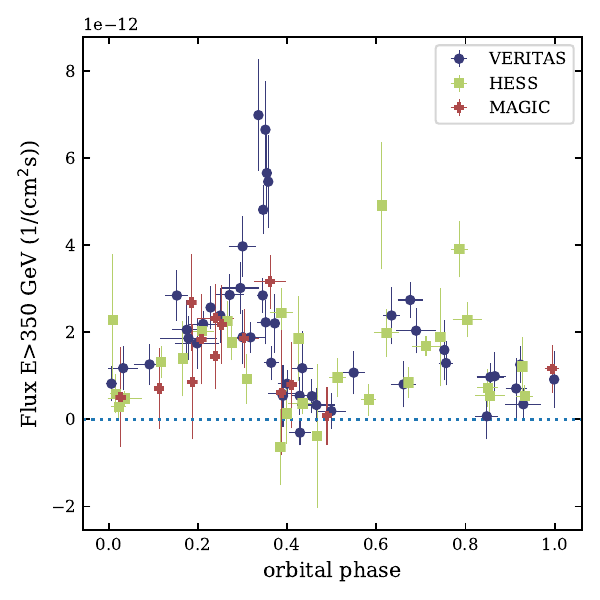}
\includegraphics[height=0.48\textwidth]{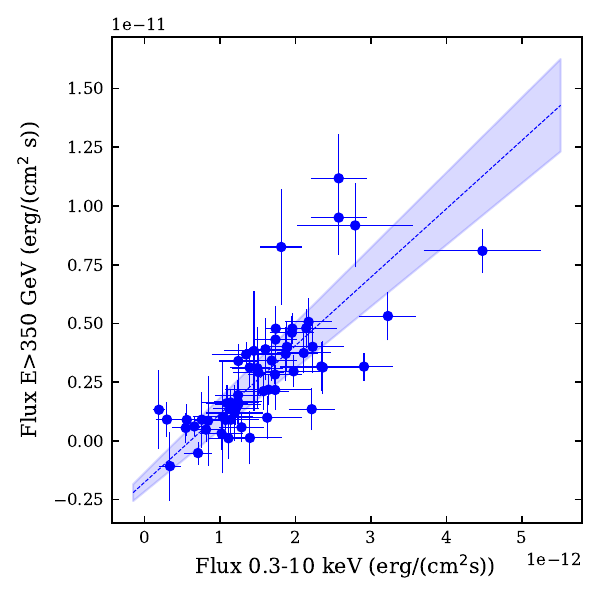}\vskip -0.1in
  \caption{ \small  (a) Left: Light curves of HESS J0632+057 measured by VERITAS, H.E.S.S. and MAGIC at energies $>350$ GeV, as a function of the orbital phase, assuming an orbital period of 317.3 days~\cite{bongiorno2011}, using about 15 years of data. (Figure from~\cite{hessJ0632_15years}). (b) Right: 
Correlation of gamma-ray ($>350$ GeV) and X-ray (0.3–10 keV) integral fluxes, measured 
contemporaneously. A fit to a straight line is shown as a dashed line, with the one sigma error range indicated with a shaded area. Details are in the original reference~\cite{hessJ0632_15years}.}
\label{fig:veritas_hessj0632}
\end{figure}

Since these early studies of HESS J0632+057, VERITAS has monitored the source extensively at energies above 350 GeV, over a 15-year period between 2004 and 2019, jointly with H.E.S.S. and MAGIC, and has recently reported on broadband analysis at X-ray energies with Swift-XRT, Chandra, XMM-Newton, NuSTAR, and
Suzaku~\cite{hessJ0632_15years}. Detailed observations were carried out for all orbital phases during this study. Figure~\ref{fig:veritas_hessj0632}a shows the gamma-ray light curve measured by VERITAS, H.E.S.S. and MAGIC as a function of orbital phase.  
The data from the three IACTs are in agreement with each other, with a maximum flux detected in the phase range 0.2-0.4, and a minimum at phases 0.4-0.6. The orbital period at TeV energies was measured to be $316.7\pm 4.4$ days, in agreement with that measured by Swift-XRT at X-ray energies~\cite{hessJ0632_15years}. 

Figure~\ref{fig:veritas_hessj0632}b shows a correlation of integral fluxes between contemporaneous gamma-ray ($>350$ GeV) and  X-ray (0.3–10 keV) data for HESS J0632+057. The gamma-ray and X-ray fluxes are found to be strongly correlated, suggesting a common origin of radiation, one that can be explained by a simple one-zone leptonic emission model,
in which relativistic electrons lose energy by synchrotron and inverse-Compton processes. These results are from the deepest study of the source with H.E.S.S.,
MAGIC and VERITAS, that consists of a total of 450 h of data~\cite{hessJ0632_15years}, and will provide an opportunity for future theoretical work to understand the causes of VHE variability, and the processes for high-energy emission from binaries. 

\begin{figure} [b!]
\centering
\vskip -0.1in
\includegraphics[height=0.38\textwidth]
 {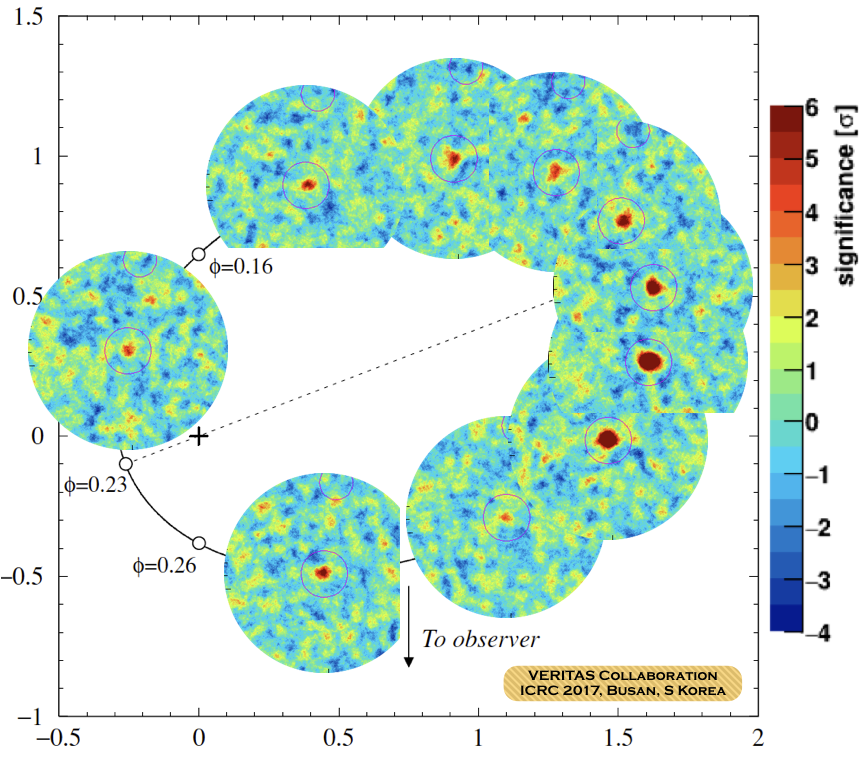}
 \includegraphics[height=0.35\textwidth]
 {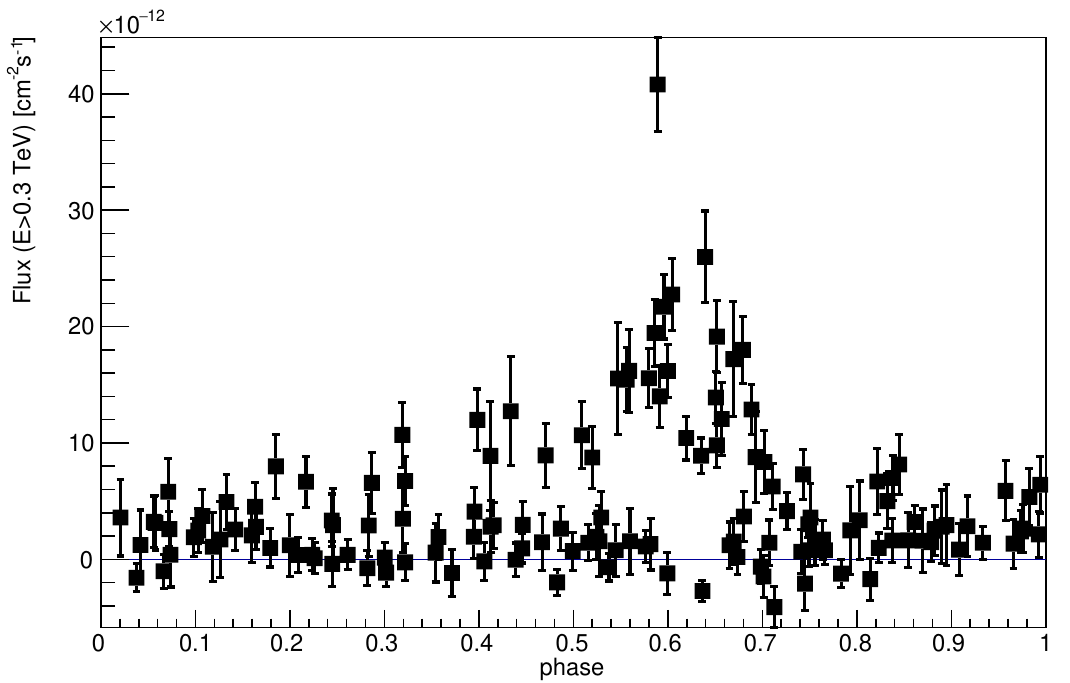}
  \caption{ \small  
  (a) Left: sky maps of \lsi\ measured by VERITAS during ten different phase bins of width $0.1$ (image locations are approximate and for illustrative purposes only), with time progression in the counter-clockwise direction. The TeV binary is detected at greater than $5\sigma$ significance level in all but one of the phase bins around the orbit. (Figure from~\cite{kar_lsi2017}.) (b) Right: Light curve of \lsi\, nightly averaged flux with an orbital phase bin of $0.1$, using more than ten years of VERITAS data. The flux peaks in the phase range 0.6-0.7, with the highest flux at a phase of 0.65.  
  (Figure from~\cite{kieda_lsi2021}.)}
\label{fig:lsi_skymap}
\end{figure}

Another binary extensively studied by VERITAS, with more than 200 hours of data, is the high-mass X-ray binary \lsi, consisting of a system with a massive B0 Ve star and an unknown compact orbiting object~\cite{kar_lsi2017, kieda_lsi2021}.  \lsi\ is variable at TeV energies, varying over the course of a single orbital cycle of 26.5 days, and has been detected across all energy bands from radio to TeV~\cite{veritas_lsi_2009}. Figure~\ref{fig:lsi_skymap}a shows sky maps of \lsi\ for ten different phase bins, generated using more than ten years of VERITAS data. Figure~\ref{fig:lsi_skymap}b shows the light curve of \lsi\ derived from more than 150 hours of data taken between 2008 and 2021. The data show a burst at an orbital phase of about 0.65~\cite{kieda_lsi2021}. VERITAS measured the peak of the gamma-ray emission in the orbital phase 0.6 to 0.7, during its apastron passage. 

\lsi\ shows orbit-to-orbit variability. The baseline TeV emission and VHE outbursts near apastron can be explained by the so-called flip-flop model, where the system changes between an accretion regime at periastron to an ejector regime at apastron, when particles are accelerated to TeV energies in the collision region between the neutron star and stellar wind~\cite{torres2012}. Super-orbital variability of about 4.5 years, reported by MAGIC and also seen in radio, X-ray and GeV bands~\cite{ahnen_lsi2016}, has not yet been seen by VERITAS, and further studies are underway. 

TeV J2032+4130 was originally detected at VHE by HEGRA~\cite{aharonian2005_hegra} but no counterpart was detected at other wavelengths, making it the first unidentified TeV source and earning it the title of "dark accelerator". In 2009 Fermi-LAT reported the identification of the pulsar PSR J2032+4127 with TeV J2032+4130~\cite{fermi_tev2032}. Further studies showed that the pulsar is in a binary system with a B0Ve star, MT91 213; the orbit was found to be highly eccentric, with a period of 20-30 years, and periastron passage was predicted to occur in early 2018~\cite{lynn2015}. Subsequent refinements suggested a period of about 50 years, with periastron passage in November 2017~\cite{Ho_PSR2032_2017}. VERITAS carried out coordinated observations with MAGIC  between 2016 and 2018, particularly during the periastron passage, which was confirmed to occur in November 2017. These observations resulted in the detection of VHE gamma-ray emission~\cite{holder_atel2017} and the discovery of a point source VER J2032+414 at the location of PSR J2032+4127~\cite{veritas_tev2032_2018}. This once-in-a-lifetime observation of the source during periastron helped in the identification of TeV J2032+4130 as a binary, one of only two TeV binaries with a pulsar as the identified compact object~\cite{veritas_tev2032_2018}. The other is the pulsar - Be-star binary system PSR B1259-63 / LS 2883~\cite{aharonian2005_ls2883}. 

\begin{figure} [t!]
\centering
\vskip -0.1in
\includegraphics[height=0.4\textwidth]
 {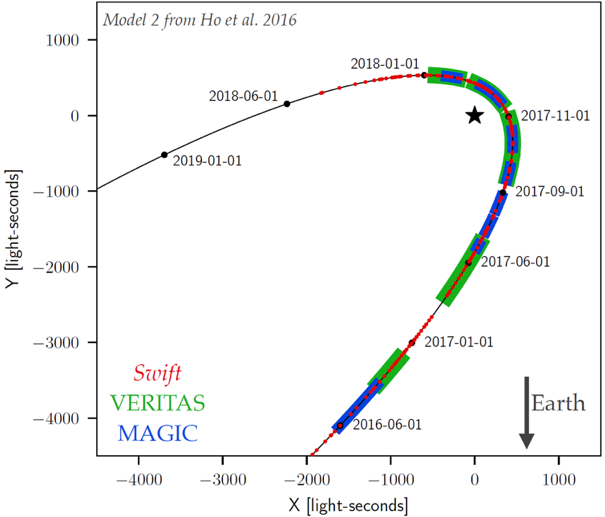}
 \includegraphics[height=0.43\textwidth]
 {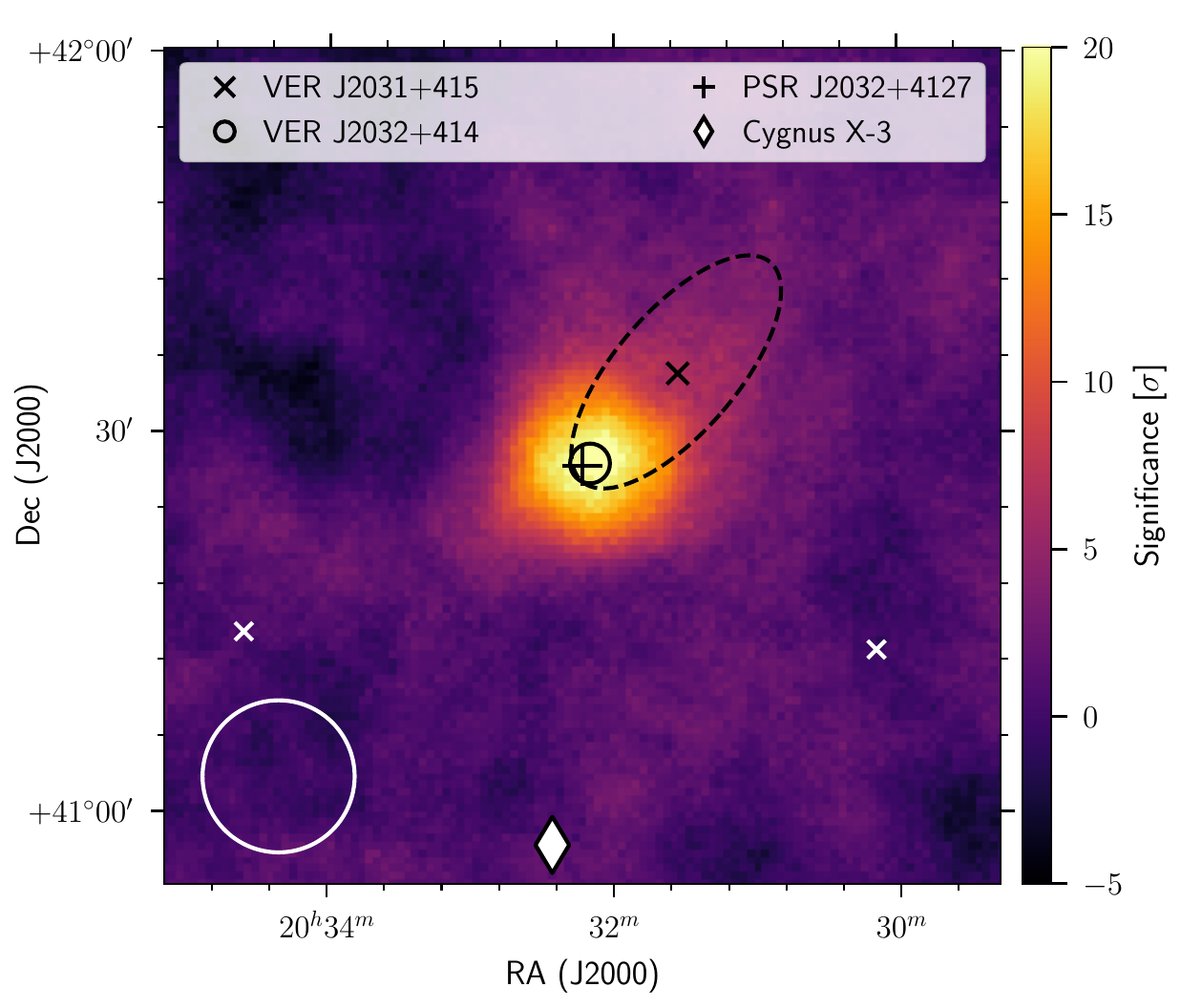}
  \caption{ \small   
 (a) Left: Schematic diagram showing the approximate orbital motion of PSR J2032+4127 around its Be star companion MT91 213 (denoted by the black star symbol), taken from~\cite{Ho_PSR2032_2017}, superimposed on which are the times of coordinated observations of the region around TeV J2032+4130 by VERITAS, MAGIC and Swift~\cite{veritas_tev2032_2018}. (b) Right: Significance sky map of the region around TeV J2032+4130, observed by VERITAS at the time of the periastron passage in November 2017~\cite{veritas_tev2032_2018}. Bright flaring activity was detected at the position of PSR J2032+4127/MT91 213, the location of which is shown as a black “+.” The dashed curve corresponds to the extension measurement of TeV 2032+4130, with the centroid of the emission measured by VERITAS shown as a black “×.” The approximate point-spread function for VERITAS at 1 TeV is shown as a white circle with radius of $0.1^{\circ}$.} 
 \label{fig:tev2032}
\end{figure}

Figure~\ref{fig:tev2032}a shows the predicted orbit of TeV J2032+4130 around periastron passage, and the times of the observations by VERITAS, MAGIC and Swift. Figure~\ref{fig:tev2032}b shows the VERITAS sky map observed in November 2017, when there was bright flaring activity from the binary~\cite{veritas_tev2032_2018}. The  VHE gamma-ray emission from this binary system is likely due to the interaction of the pulsar wind with the wind and/or disk of the Be star. 

\subsection{Multimessenger Partnership}

\begin{figure} [t!]
\centering
\vskip -0.1in
\includegraphics[height=0.5\textwidth]
 {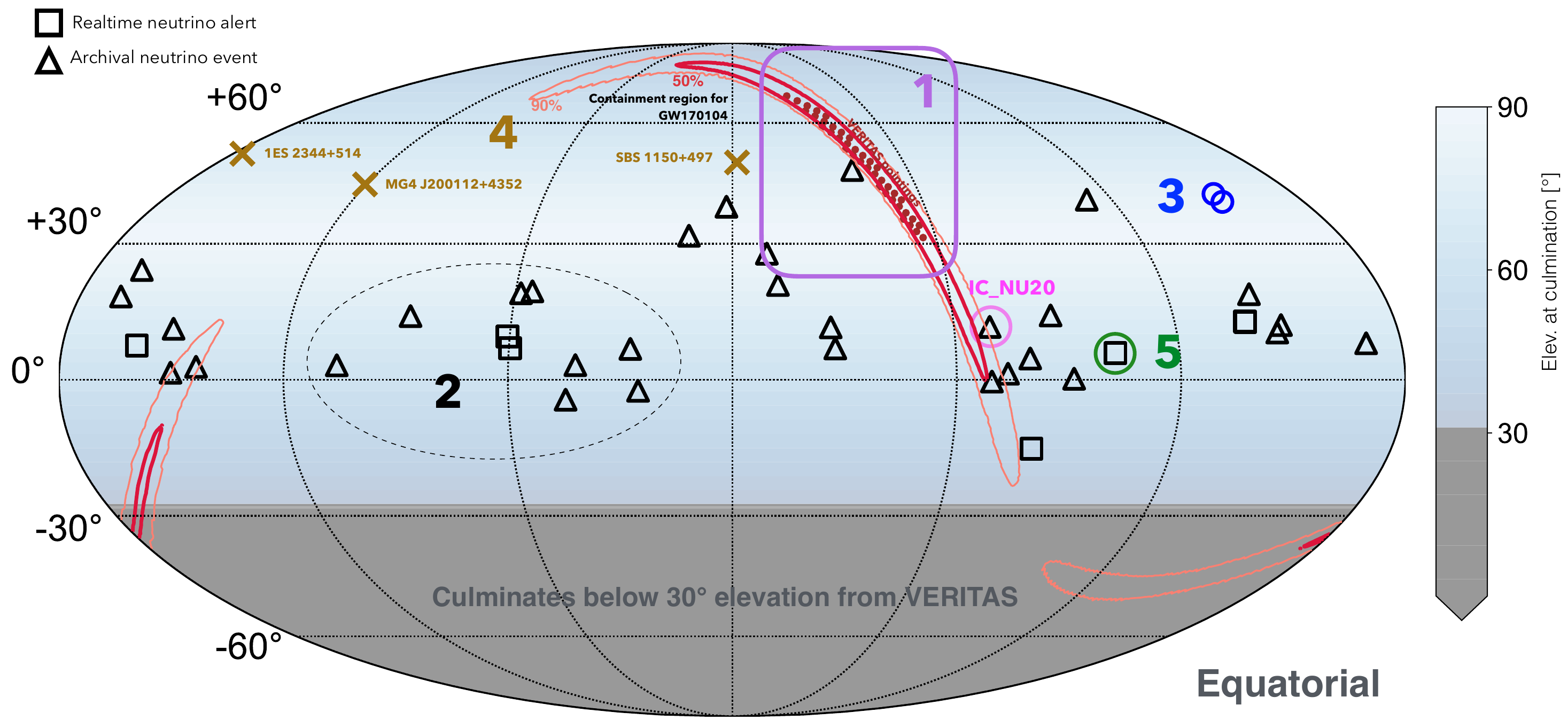}\vskip -0.1in
  \caption{ \small 
Sky map showing multimessenger triggers followed up by VERITAS. The background color shows the the elevation angle at culmination from the VERITAS site for different declination bands. The numbers
correspond to the following observations: GW events (1), realtime and archival single-neutrino events from IceCube (2), IceCube neutrino multiplets (3), potential neutrino emission from known gamma-ray sources (4) and long-term monitoring of the candidate neutrino blazar TXS 0506+056 (5). The squares represent realtime neutrino alerts and the triangles represent archival neutrino events. (Figure from~\cite{santander_2019ICRC_MM}.) }
\label{fig:mm_veritas}
\end{figure}

The VERITAS multimessenger program consists of follow-up observations of candidate high-energy (HE) neutrino detections announced by IceCube and gravitational-wave (GW) events from LIGO/Virgo. Additionally, VERITAS partners with LHAASO, HAWC, Fermi-LAT and Swift, as well as other instruments for following up on transient events and ``ToO" or target-of-opportunity alerts. VERITAS has carried out observations of ``hotspots'' in the IceCube point-source map, clusters of neutrino events around known VHE sources, IceCube archival neutrino events, and, starting in 2016, real-time alert follow-ups of IceCube neutrinos~\cite{veritas_icecube} (see Fig.~\ref{fig:mm_veritas} for a sky map of multimessenger triggers observed by VERITAS). 

In 2018, IceCube reported the detection of the HE neutrino source IC~170922A, spatially and temporally coincident with the flaring blazar TXS 0506+056~\cite{IC170922A}. A  gamma-ray flare observed by Fermi-LAT~\cite{IC170922A_Fermi} was followed by VHE detections of the source by MAGIC~\cite{IC170922A_MAGIC} and soon afterwards by VERITAS~\cite{IC170922A_VERITAS}. Although this was not conclusive proof that TXS 0506+056 is an extragalactic HE neutrino source, astrophysical neutrinos are expected to originate from hadronic interactions in or near cosmic-ray accelerators, 
and spatial and temporal correlations between neutrino events and gamma rays are expected~\cite{resconi2017}. Gamma-ray observatories are key in identifying the sources of astrophysical neutrinos, and are hence the sites of hadronic cosmic ray acceleration at the highest energies. Given that IACTs have the best sensitivities for observing gamma-ray emission on short timescales, and excellent angular resolution at gamma-ray energies, ToO observations by IACTs to follow-up  multimessenger sources are essential.

Stellar core collapses and compact binary mergers can result in both gravitational radiation and electromagnetic emission at high energies, as was demonstrated by the detection of GW~170817~\cite{LIGO_GW170817} by LIGO and gamma rays from the event by Fermi-LAT~\cite{Fermi_GW170817}. VERITAS has searched for electromagnetic emission from 12 gravitational-wave (GW) sources in the third observing run of LIGO,  O3, resulting in no detections~\cite{santander_2019ICRC_MM}. In addition, an interesting study that was pioneered by VERITAS was the serendipitous search for VHE emission coincident with subthreshold GW binary neutron star (BNS) merger candidates identified in Advanced LIGO’s first observing run (O1)~\cite{subthresh_veritas_2021}. The goal was to explore whether ``subthreshold''  candidates, gravitational-wave events in the LIGO/Virgo data with low signal-to-noise ratio that do not qualify as significant candidates, 
together with contemporaneous gamma-ray observations could lead to significant detections. Eight hours of archival VERITAS observations coincident with seven BNS merger candidates were identified for this search. 
No VHE emission was detected, but the measured integral flux upper limits for the coincident regions were used to calculate the fluence which was then compared to a short GRB afterglow model, extrapolated from a representative burst GRB 090510 to higher energies and longer durations. The VERITAS upper limits fall
orders of magnitude below the predictions of this model. This suggests that any synchrotron emission from the
forward shock, if it extends up to the VHE range, was not observed by VERITAS, or that the event was much less luminous.

GRBs are the brightest gamma-ray transients but evidence for VHE emission remained elusive until recently, when H.E.S.S. and MAGIC both reported the detection of VHE gamma rays from GRBs (see~\cite{magic_GRB_2019}, for example). VERITAS has yet to detect TeV emission from a GRB, despite rapid responses to a large number of GRB alerts.  GRBs continue to be one of the highest priority targets for VERITAS~\cite{veritas_grb_weiner_2018}. 

Another high-priority program for VERITAS is the search for VHE emission from fast radio bursts (FRBs), energetic bursts of radio emission on ms timescales. These sources are extragalactic, and their nature is uncertain, but in some cases they have been associated with magnetars~\cite{FRB_magnetars}. VERITAS has an ongoing program of coordinated observations with the CHIME FRB collaboration~\cite{chime}, whose telescopes are fortuitously located only 8.6 degrees west and 17.8 degrees north of VERITAS, allowing for the possibility of simultaneous observations~\cite{lundy_2021}. Given the unpredictable and episodic nature of the burst emission, simultaneous observations are critical.

\subsection{Using VERITAS Data to Explore Fundamental Physics}

The large data set that VERITAS has acquired since first light naturally lends itself to archival analysis and fundamental physics studies, particularly where signals are expected to be weak. Some of the topics explored need no special targeted observations, and can simply take advantage of the vast amount of data collected during astrophysical observations. 

\begin{figure} [t!]
\centering
\vskip -0.1in
\includegraphics[height=0.41\textwidth]{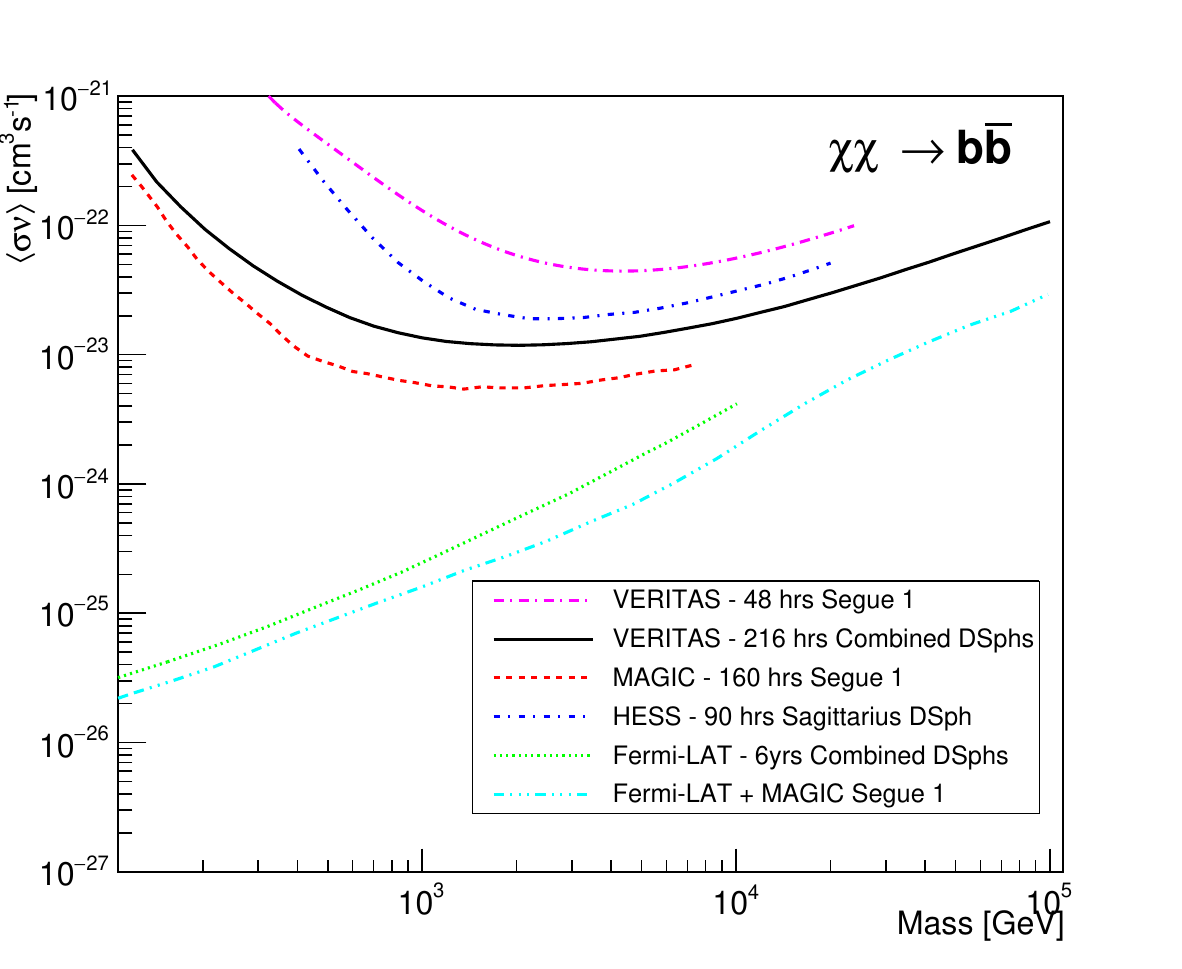}\hskip -0.2in
%Fig. 9 left: https://arxiv.org/pdf/1703.04937.pdf
\includegraphics[height=0.415\textwidth] {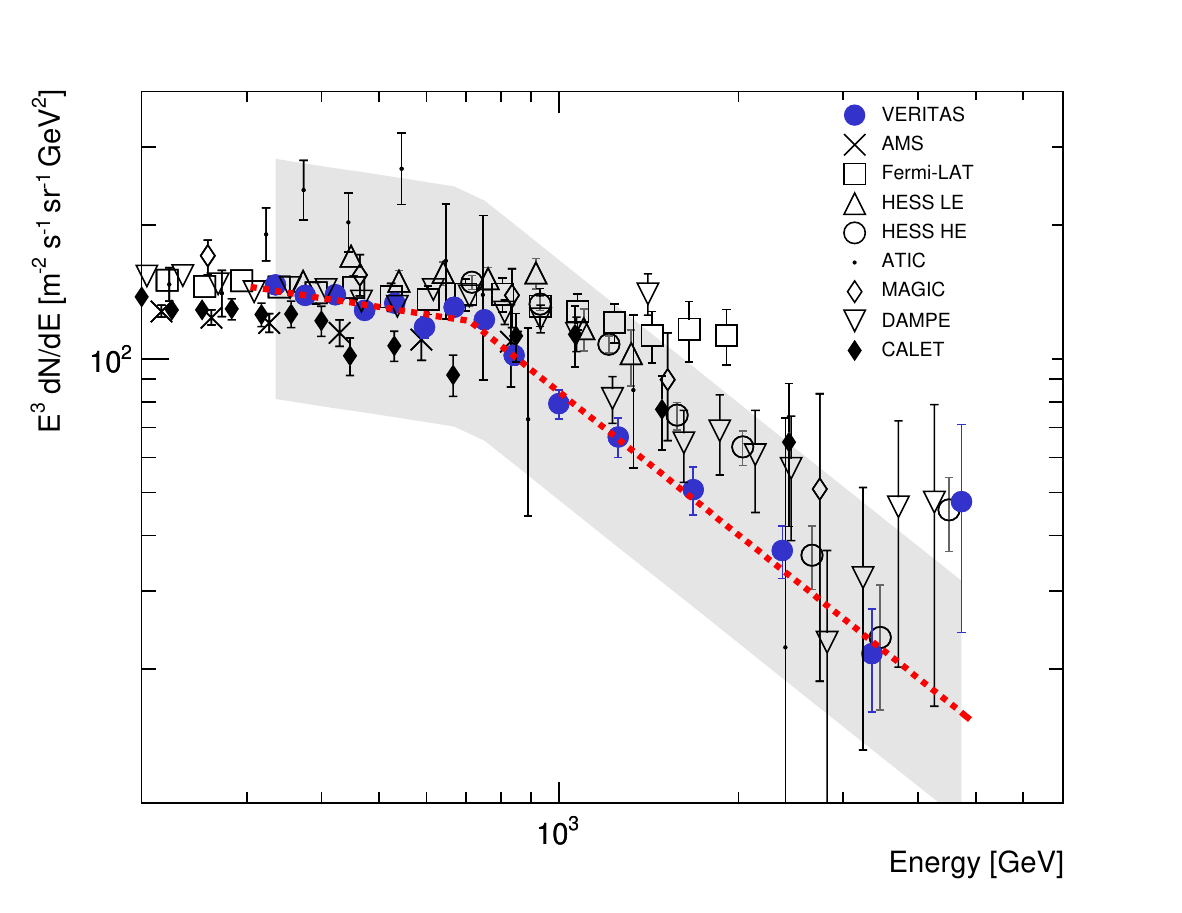}
%Fig. 2: https://arxiv.org/pdf/1808.10028.pdf
\vskip -0.1in
  \caption{ \small 
  (a) Left: Limits on the cross section of annihilation for dark matter into the bottom quark channel for a combined analysis on several dwarf spheroidal galaxies observed by VERITAS. Results from several other analyses are also shown. Figure from~\cite{veritas_dwarf_2017}. (b) Right: Spectrum of cosmic ray electrons measured by VERITAS in the energy range 300 GeV and 5 TeV. Historical measurements are also shown. (Figure and details in ~\cite{veritas_cre_2018}.) }
\label{fig_veritas_dwarf_archambault2017}
\end{figure}

VERITAS has presented competitive limits on the annihilation cross section of weakly interacting massive particles (WIMPs), candidates for dark matter, by carrying out a combined analysis on about 230 h of data on four dwarf galaxies: Segue1, Ursa Minor, Draco, Bo\"otes. No evidence of any gamma-ray emission
was detected either in the combined data set or from any individual dwarf galaxy. Upper limits at 1 TeV were derived for dark matter annihilating into various possible final states: bottom quark ($b\bar b$), tau lepton ($\tau^+\tau^-$), and gauge-boson ($\gamma\gamma$)~\cite{veritas_dwarf_2017}. Figure~\ref{fig_veritas_dwarf_archambault2017}a shows the limits derived for annihilation into the bottom quark channel, together with those obtained from other studies. The VERITAS analysis resulted in lower limits about two orders of magnitude above the
relic abundance value of $10^{-26}$ cm$^{-2}$ of the annihilation cross section. 
A joint analysis including data from Fermi-LAT and from other IACTs could lead to more constraining limits.

Archival data from VERITAS may be used to constrain the evaporation rate of primordial black holes (PBHs) by searching for bursts of gamma-ray emission in the data, predicted in certain models. PBHs may have been formed as a result of density fluctuations in the very early Universe from the gravitational collapse of over-dense regions. A search was carried out using 700 hours of VERITAS data for
evidence of burst-like emission due to evaporating PBHs and a constraint was placed on the number density of PBHs~\cite{tesic2012}.

VERITAS archival data provided an opportunity to indirectly search for cosmic ray electrons and positrons (CREs) and to measure their energy spectrum. 
Figure~\ref{fig_veritas_dwarf_archambault2017}b shows the CRE spectrum measured by VERITAS in the energy range from 300 GeV to about 5 TeV~\cite{veritas_cre_2018}, which goes beyond the energy range explored by Fermi-LAT and AMS-02. The VERITAS spectrum shown in the figure cannot be fitted by a single power-law function, and a broken power-law function, with a break energy at $710 \pm 40_{\rm stat}\pm 140_{\rm syst}$ GeV is needed to explain the data. A break in the spectrum at $\sim 1$ TeV is also seen by H.E.S.S. and the space-borne cosmic ray experiments, CALET and DAMPE, but has not been detected by MAGIC. 

Since CREs suffer rapid energy losses due to synchrotron radiation and inverse-Compton
scattering while propagating within the Galaxy, they can be used to probe only the local Galactic neighborhood, which in the case of electrons with TeV energies is less than about 1 kiloparsec. Sources accelerating electrons to TeV energies could be supernova remnants and pulsars, such as Geminga and Monogem in the Milky Way Galaxy. CREs may also be produced in the annihilation or decay of heavy dark matter particles. 

VERITAS has also reported the measurement of the spectrum of cosmic ray iron nuclei in the energy range from 20 TeV to 500 TeV. Iron is the third most abundant element in the composition of cosmic rays at TeV energies. A template-based analysis method was used for the first time for energy reconstruction of iron-induced air showers~\cite{veritas_cosmicray_iron_2018} in the VERITAS data. Development and further enhancements of image-template analyses may prove useful in measuring the spectra of other cosmic ray elements in the future. 

\section{Legacy and Prospects for the Future}

So far the VERITAS collaboration has published more than 150 papers, mostly peer-reviewed articles in journals but also including conference proceedings. These are the lasting contributions to knowledge that have been made by the group. To enable members of the community to use these findings effectively, high-level data products from all publications are being made available through the online github repository in the VERITAS data catalog VTSCat~\cite{VTSCat}. The products include spectral flux points, light curves, spectral fits, and sky maps. Data from published tables, such as upper limits from dark matter searches, are also available. Although some measurements will likely become less relevant as future, more-sensitive observatories provide more precise numbers, many, such as flares will have lasting value, due to the dynamic and episodic nature of astrophysics.

More than 70 people have earned their doctoral degrees while doing research with VERITAS. Some have stayed in gamma-ray astronomy, moving to other collaborations, like Fermi-LAT or the Cherenkov Telescope Array (CTA)~\cite{cta}, or working as postdoctoral researchers at a different institution within VERITAS. Others have migrated to other projects in particle astrophysics, such as IceCube, or gone into accelerator-based particle physics. The rest have pursued careers in teaching or industry. The same can be said of the more than 80 postdoctoral researchers that have spent time in the collaboration. This human capital is an important part of the VERITAS legacy.

VERITAS will continue to be run for the foreseeable future. Running a detector that is more than 15 years old has its challenges, but these are being met and most of the instrument is in remarkably good condition.  However, when CTA is up and running, with its superior sensitivity it will be harder to motivate a regular observing program with VERITAS, although target-of-opportunity observations in response to multimessenger alerts, and tailored monitoring programs, may still be valuable, at least in the initial CTA era.

VERITAS is already being used to help commission a new telescope, the prototype Schwarzschild-Couder Telescope (pSCT)~\cite{pSCT}, located near the array. It is seen as a possible contribution of US groups to the CTA project. Aside from gamma-ray astronomy, the collaboration is taking advantage of the large mirrors of the four telescopes for other applications that may continue farther into the future. Measurements of stellar diameters using intensity interferometry~\cite{sii-2}, and exploration of Kuiper-belt objects and exoplanets using stellar occultation, are two such activities. Members of the collaboration, supported in part by the Breakthrough Listen Initiative, are active in the optical search for extraterrestrial intelligence (OSETI), looking for laser pulses from distant civilizations~\cite{seti-1}. 
This additional application of imaging atmospheric-Cherenkov telescope arrays, to search for extremely short duration optical transients, could potentially impact future telescope designs.

\bigskip
\noindent
VERITAS research is supported by grants from the U.S. DoE, the NSF, and the Smithsonian, by NSERC in Canada, and by the Helmholtz Association in Germany. RM acknowledges support from NSF grant PHY-2110497. Hanna and Mukherjee are members of the VERITAS Collaboration (veritas.sao.arizona.edu) and acknowledge the support of their colleagues. We also gratefully acknowledge the excellent work of VERITAS technical support staff.

\end{document}